\documentclass[conference]{IEEEtran}

\usepackage[T1]{fontenc}
\usepackage{graphicx}
\usepackage{float}
\usepackage{comment}
\usepackage{subcaption}
\usepackage{textcomp}
\usepackage{xcolor}
\usepackage{hyperref}
\usepackage{enumitem}
\usepackage{listings}
\usepackage{amsmath}

\usepackage{tikz}

\newcommand\copyrighttext{%
  \fontsize{7pt}{7pt}\selectfont \textcopyright 2026 IEEE. Personal use of this material is permitted.
  Permission from IEEE must be obtained for all other uses, in any current or future
  media, including reprinting/republishing this material for advertising or promotional
  purposes, creating new collective works, for resale or redistribution to servers or
  lists, or reuse of any copyrighted component of this work in other works. \\ \\
  M. Adams and A. Bienz, "Optimizing Allreduce Operations for Modern Heterogeneous Architectures with Multiple Processes per GPU," 2026 IEEE International Parallel and Distributed Processing Symposium Workshops (IPDPSW), New Orleans, LA, USA, 2026, pp. 135-144, doi: 10.1109/IPDPSW71298.2026.00023. This article can be found on IEEE \textit{Xplore} at \url{https://ieeexplore.ieee.org/document/11651997}.
  }
\newcommand\copyrightnotice{%
\begin{tikzpicture}[remember picture,overlay]
\node[anchor=south,yshift=10pt] at (current page.south) 
  {\fbox{\parbox{\dimexpr\textwidth-\fboxsep-\fboxrule\relax}{\copyrighttext}}};
\end{tikzpicture}%
}

\usepackage{setspace}

\usepackage[algo2e,ruled,vlined]{algorithm2e}
\SetCommentSty{scriptsize}
\SetKwComment{tcc}{\{}{\}}
\SetKwIF{If}{ElseIf}{Else}{if}{}{else if}{else}{endif}
\SetKwFor{While}{while}{}{endw}
\SetKwFor{ForEach}{for each}{}{endfch}
\SetKwRepeat{Do}{do}{while}
\SetAlgorithmName{Algorithm}{Algorithm}{List of Algorithms}

\graphicspath{{Figures/}}

\begin{document}

\title{Optimizing Allreduce Operations for Modern Heterogeneous Architectures with Multiple Processes per GPU}

\author{
\IEEEauthorblockN{Michael Adams}
\IEEEauthorblockA{
University of New Mexico\\
Albuquerque, NM, USA\\
mikethebos@unm.edu
}
\and
\IEEEauthorblockN{Amanda Bienz}
\IEEEauthorblockA{
University of New Mexico\\
Albuquerque, NM, USA\\
bienz@unm.edu
}
}

\maketitle
\copyrightnotice %for IEEE
\begin{abstract}
Large inter-GPU all-reduce operations, prevalent throughout deep learning, are bottlenecked by communication costs. Emerging heterogeneous architectures are comprised of complex nodes, often containing $4$ GPUs and dozens to hundreds of CPU cores per node. Parallel applications are typically accelerated on the available GPUs, using only a single CPU core per GPU while the remaining cores sit idle. This paper presents novel optimizations to large GPU-aware all-reduce operations by extending the lane-aware algorithm to heterogeneous architectures and notably using multiple CPU cores per GPU to accelerate these operations. Using GPUDirect RDMA and host copy communications respectively, these multi-CPU-accelerated GPU-aware all-reduces yield speedups over system MPI of up to $3$x on LLNL's Tuolumne supercomputer and up to $2.45$x for large MPI all-reduces across the NVIDIA A100 GPUs of NCSA's Delta supercomputer.
\end{abstract}

\begin{IEEEkeywords}
Allreduce, heterogeneous architectures, GPU, MPI, multi-lane, node-aware
\end{IEEEkeywords}

\section{Introduction}
\begin{comment}
\begin{itemize}
    \item Emerging exascale systems - heterogeneous, computation accelerated on GPUs and communication between data in gpu memory
    \item Large allreduce operations bottleneck neural networks / large language models
    \item Poor scalabilty : show plot of just system MPI GPU-Aware allreduce cost at different process counts
    \item Communication cost is dependent on locality of sending and receiving processes (on node vs off node).
    \item Communication is optimized when many processes each send an equal and minimal portion of data through the network (this is what the multi-lane paper shows)
    \item Contributions of paper: present multiple novel algorithms that use locality-awareness to reduce the amount of off node data.  Present performance results on x GPUs or nodes of a Grace Hopper system (or brief description of various systems you have results for).  Achieve x speedup over existing approaches.  Compare against NCCL and achieve up to x speedup over NCCL.  
    \item Outline of remainder of paper (e.g. in Section 2 we present...)
\end{itemize}
\end{comment}

Emerging exascale systems are comprised of heterogeneous nodes, each typically containing $4-8$ GPUs and dozens to hundreds of CPU cores.  Parallel applications achieve high performance through device utilization, accelerating computation on the available GPUs and communicating directly between devices with GPU-aware MPI.  Applications typically use a single MPI process per GPU, utilizing one core optimally located near each GPU.  While these architectures are equipped with dozens of CPU cores per node, the vast majority of these cores are unused by parallel applications.

Large all-reduce operations dominate the cost of many popular applications, including moderately sized neural networks and large language models (LLMs), in which the layers are distributed across GPUs.  
\begin{figure}[ht!]
    \centering
    \includegraphics[width=\linewidth]{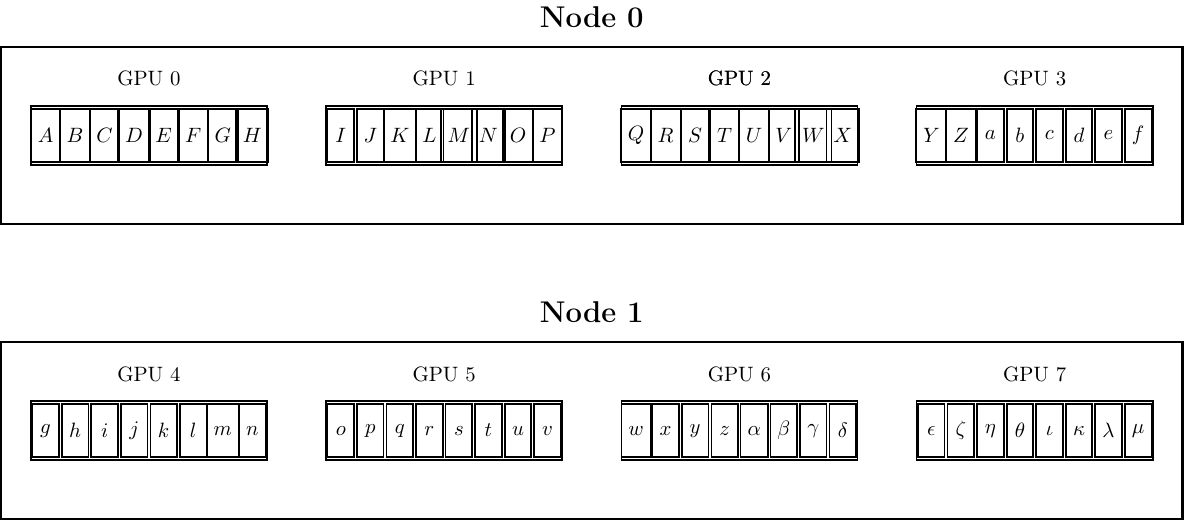}
    \caption{Example All-Reduce Setup}
    %\Description{Example buffers to be reduced across $8$ GPUs and $2$ nodes.}
    \label{fig:allreduce_ex}
\end{figure}
These reductions require each process to exchange and reduce data with all other processes.  For example, Figure~\ref{fig:allreduce_ex} shows input buffers across $8$ GPUs that are to be reduced.

\begin{figure}[ht!]
    \centering
    \includegraphics[width=\linewidth]{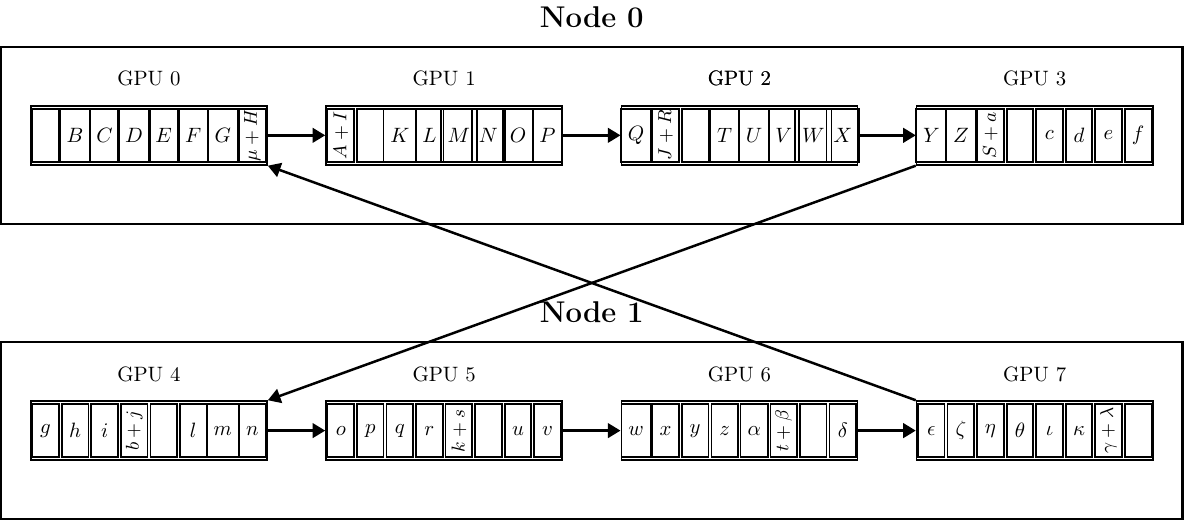}
    \caption{Ring All-Reduce, First Step\vspace{-1em}}
    %\Description{The first step of a ring all-reduce.}
    \label{fig:allreduce_ring}
\end{figure}

While there are many existing all-reduce algorithms, the standard approach for large data sizes is through a ring pattern~\cite{ring0}, in which every process exchanges a portion of the data with neighbors at each step, the first of which is displayed in Figure~\ref{fig:allreduce_ring}.
The cost of each message varies with the relative locations of the sending and receiving processes~\cite{perf_model_locality}, and the ring algorithm attempts to optimize locality by restricting communication to neighboring ranks.
%The ring algorithm exhibits limited scalability as either data size or process count increases.  To reduce a buffer of $n$ bytes among $p$ processes with a ring algorithm, $2*(p-1)$ messages are communicated, with each message containing $\frac{n}{p}$ bytes.  In the ring algorithm, each process $r$ communicates only with neighboring ranks $(p+r-1)\mod{p}$ and $(r+1)\mod{p}$.  
Assuming an optimal process layout, the majority of processes may communicate locally, such as within a node.  For example, the majority of messages displayed in Figure~\ref{fig:allreduce_ring} are intra-node.  However, at the boundary, processes must communicate with neighboring nodes at each step.  

\begin{figure}[ht!]
    \centering
    \includegraphics[width=\linewidth]{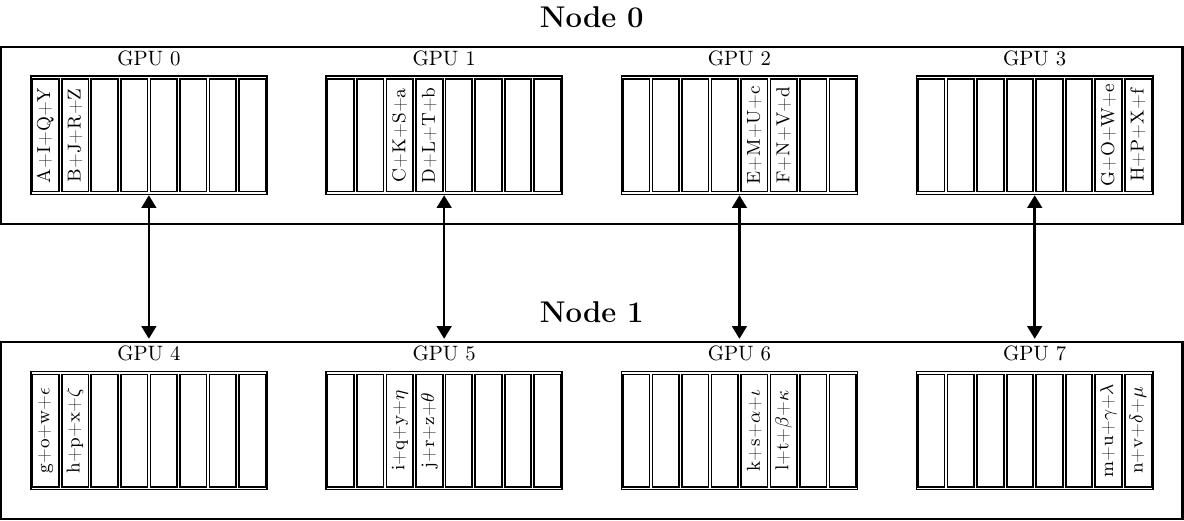}
    \caption{Multi-Lane All-Reduce, Inter-Node Step\vspace*{-4mm}}
    %\Description{The inter-node step of a multi-lane all-reduce.}
    \label{fig:allreduce_lane}
\end{figure}

The multi-lane all-reduce~\cite{lane} improves the cost of large reductions in a topology-agnostic manner through lane-awareness, with each process per node performing a separate portion of the inter-node all-reduce, exemplified in Figure~\ref{fig:allreduce_lane}.
All processes first perform a reduce-scatter on-node, evenly partitioning the buffers before inter-node communication.  Each process then exchanges smaller buffers with the corresponding processes on each other node.  Finally, data is gathered within each node.  While more data is communicated than with the ring algorithm, the additional overheads are within each node.  

\begin{figure}[ht!]
    \centering
    \includegraphics[page=3,width=1.0\linewidth]{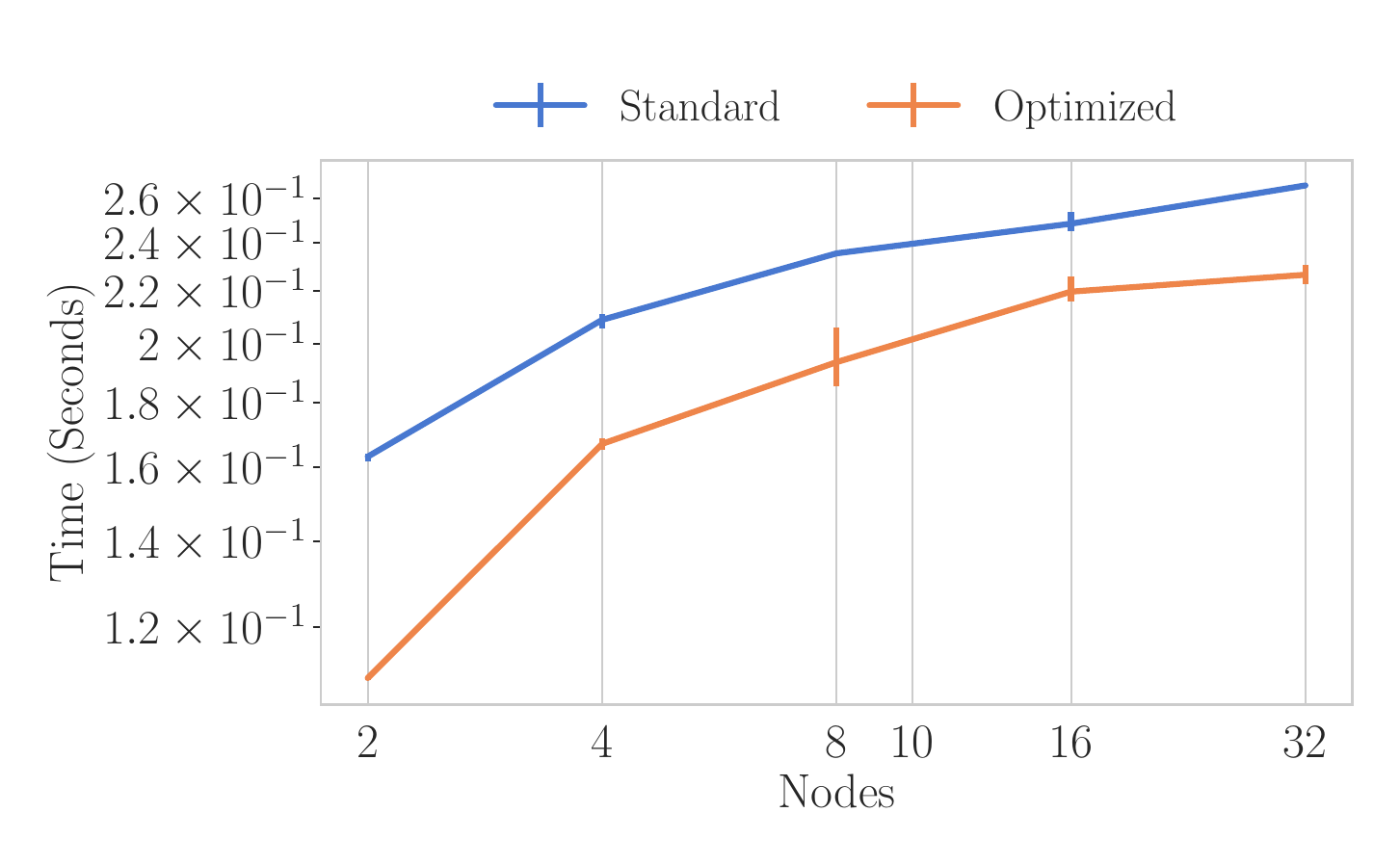} % Replace with your image file
    \caption{Standard \texttt{MPI\_Allreduce} vs our optimized all-reduce on LLNL's Tuolumne}
    %\Description{}
    \label{fig:speedup}
\end{figure}

This paper explores methods for optimizing large GPU-aware MPI collective operations for emerging heterogeneous architectures using the all-reduce as a proof-of-concept.  This paper presents a novel extension of the multi-lane algorithm to GPU-aware all-reduces backed by host copy and GPUDirect Remote Direct Memory Access (RDMA) communication on NVIDIA Ampere-based and AMD MI300A-based systems, respectively.  We further optimize the all-reduce through improved resource utilization, with multiple CPU cores each progressing a portion of the all-reduce asynchronously with performance as seen in Figure~\ref{fig:speedup}.  Finally, we analyze the performance of the all-reduces in the various partitioning modes of the AMD MI300A, showing that the multi-lane all-reduces benefit from increased logical GPU count.

This paper provides the following novel contributions:
\begin{itemize}
    \item Compares the performance of GPU-aware standard and multi-lane MPI algorithms on modern architectures, using the all-reduce as a motivating example;
    \item Uses multiple CPU cores per GPU, each progressing a portion of the all-reduce asynchronously;
    \item Presents a performance analysis of these optimizations on modern architectures, achieving up to $2.45$x speedup over OpenMPI's GPU-aware all-reduce using host copy communication and up to $1.17$x speedup over Cray MPICH's GPUDirect RDMA implementation; and
    \item Accelerates all-reduces using the TPX and CPX partitioning modes using the multi-lane algorithm, achieving up to $3$x speedup over the Cray MPICH implementation.
\end{itemize}

%Emerging exascale systems rely on heterogeneous acceleration to achieve optimal performance where communication and computation occur in graphics processing unit (GPU) memory. An example application is deep learning (DL), an universal function approximation tool used to solve scientific problems where large amounts of known data is available. These backpropagation and linear algebra operations are easily vectorized on a GPU.

The remainder of our paper is outlined as follows.  Section~\ref{sec:background} describes background information and related works, detailing the baseline ring and multi-lane all-reduce algorithms. Section~\ref{sec:microbenchmarks} provides benchmarks of point-to-point communication using multiple processes per GPU to motivate our work. The methodology for optimizing large all-reduces is described in Section~\ref{sec:methods}.  Section~\ref{sec:results} provides performance results for our approaches, with results using host copy communication shown in Section~\ref{sec:host_copy_results} and extensions to GPUDirect RDMA communication on the AMD MI300A shown in Section~\ref{sec:gpudirect_results}. Performance on the various partitioning modes of the AMD MI300A is shown in Section~\ref{sec:mi300a_partitioning}.  Finally, Section~\ref{sec:conclusions} provides concluding remarks.
 
%The following sections in our paper are outlined below. In Section 2, we present existing approaches in node-aware, lane-aware, and topology-aware collective communication, as well as the normal algorithms used in GPU-aware all-reduces. In Section 3, we provide pseudocode and running time analysis for our standard and lane-aware all-reduce algorithms. In Section 4, we show speedup in performance for large arrays on Delta, Tioga, and Tuolumne. In Section 5, we discuss the application of our optimizations to all GPU-aware collective communication patterns based on macro-scale topologies of systems and their emerging integrated memory architectures.

\section{Background}\label{sec:background}

All-reduce operations are generally performed using three different algorithms:  recursive doubling~\cite{recursive_doubling}, Rabenseifner's algorithm~\cite{ring0}, and ring~\cite{ring0}.  Recursive doubling is used for small reductions, minimizing the number of messages to $\log{p}$ where $p$ is the number of processes, but sending all $n$ bytes of the buffer in each message.  
%each rank $r$ reduces with process $r \oplus 2^s$ where $s$ is the number of messages sent so far.  
Rabenseifner's algorithm improves performance for moderately-sized buffers, with each process reducing a fraction of the buffer before gathering all reductions.  The total number of messages increases to $2\cdot \log{p}$, but the amount of data communicated and reduced upon is cut in half at each step.
%For larger messages, a Reduce=scatter Allgatherv algorithm is used to minimize the message size by $n$; each rank $r$ accumulates the $r*(size/n)$th chunk of data of all processes during the Reduce-scatter phase then all chunks are sent to all other processes during the Allgatherv phase. 
The ring all-reduce further optimizes for large reductions by restricting communication only to immediate neighbors, increasing the message count to $2\cdot (p-1)$~\cite{ring0,ring1}.  In recent years, this algorithm has been adopted within deep learning applications~\cite{ring2,ring3}.
%With the advent of deep neural networks yielding even larger buffers, the ring all-reduce (\autoref{alg:ring}) uses the same structure of the latter algorithm, but, in each phase, rank $r$ communicates with the same 2 neighboring processes $n - 1$ times; assuming neighboring ranks are topologically close, this reduces the transmission time of each message.

% MPI_Neighbor_alltoallv_init(
%    const void *sendbuf, 
%    const int sendcounts[], 
%    const int sdispls[], 
%    MPI_Datatype sendtype, 
%    void *recvbuf, 
%    const int recvcounts[], 
%    const int rdispls[], 
%    MPI_Datatype recvtype, 
%    MPI_Comm comm,
%    MPI_Info info,
%    MPI_Request *request)

\begin{algorithm2e}[ht!]
  \DontPrintSemicolon%
  \KwIn{$\texttt{rank, count, size from communicator}$}

  \BlankLine%
    Get $r$, $c_{buf}$, $n$ from communicator\;
    \BlankLine%

    $c_{chunk} = c_{buf} / n$

    $c_{on chunk}[n]$

    \For{$i \gets 0$ \KwTo $n$}{
        $c_{on chunk}[i] = c_{chunk}$
    }
    \BlankLine%

    \tcp{displacements}
    $D[n]$
    $D[0] = 0$
    
     \For{$i \gets 1$ \KwTo $n$}{ 
         $D[i] = D[i - 1] + c_{on chunk}[i - 1]$
     }

    $sp = r$, $rp = r - 1$ if $r \neq 0$ else $n - 1$

    $buf_{sendfrom} = buf_{send}$
    
    \For{$i \gets 0$ \KwTo $n - 1$}{
        \tcp{order based on rank}
        MPI\_Reduce to next process in ring of $c_{on chunk}[sp]$ counts of $buf_{sendfrom}$ at offset $D[sp]$
        
        MPI\_Reduce from previous process in ring of $c_{on chunk}[rp]$ counts of $buf_{send}$ at offset $D[rp]$ to $buf_{recv}$ at offset $D[rp]$

        $sp = sp - 1$ if $sp \neq 0$ else $n - 1$
        
        $rp = rp - 1$ if $rp \neq 0$ else $n - 1$

        $buf_{sendfrom} = buf_{recv}$
    }

    $sp = r + 1$ if $r \neq n - 1$ else $0$, $rp = r$
    
    \For{$i \gets 0$ \KwTo $n - 1$}{
        MPI\_Isend to next process in ring of $c_{on chunk}[sp]$ counts of $buf_{recv}$ at offset $D[sp]$
        
        MPI\_Irecv from previous process in ring of $c_{on chunk}[rp]$ counts of $buf_{recv}$ at offset $D[rp]$

        MPI\_Waitall

        $sp = sp - 1$ if $sp \neq 0$ else $n - 1$
        
        $rp = rp - 1$ if $rp \neq 0$ else $n - 1$
    }
  
    \caption{\texttt{ring\_allreduce}}\label{alg:ring}
\end{algorithm2e}

The ring implementation for large all-reduce operations, detailed in Algorithm~\ref{alg:ring}, requires each process to send a fraction of the buffer to one neighbor $n-1$ times, while receiving a separate fraction of the buffer from its other neighbor.  The received message is then reduced with the local buffer, before the process is repeated with the next successive fractions of the buffer.  Finally, in a similar manner, these fractions are gathered to all ranks by communicating with a process's neighbors.

While the ring algorithm improves the locality of communication over other tree-based approaches, there remains a large variability among these neighboring messages.  Assuming there are $4$ processes per node, two processes on each node will communicate only intra-node messages, while the other two will either send to or receive from a neighboring node.  
%Figure~\ref{TODO} shows the cost of sending a single inter-GPU message of various sizes on Lawrence Livermore National Laboratory's Tuolumne.  For large messages, the performance of inter-node and intra-node messages vary significantly.

% MPI_Neighbor_alltoallv_init(
%    const void *sendbuf, 
%    const int sendcounts[], 
%    const int sdispls[], 
%    MPI_Datatype sendtype, 
%    void *recvbuf, 
%    const int recvcounts[], 
%    const int rdispls[], 
%    MPI_Datatype recvtype, 
%    MPI_Comm comm,
%    MPI_Info info,
%    MPI_Request *request)

\begin{algorithm2e}[ht!]
  \DontPrintSemicolon%
  \KwIn{$\texttt{rank, count,}$\\
  $\texttt{\quad \quad \quad size from group communicator}$}
%  \KwIn{$\texttt{size from group communicator}$}

  \BlankLine%

    Get $r$, $c_{buf}$, $n_{group}$ from communicator\;
    $comm_{group}$\;
    $comm_{lane}$\;

    $c_{group} = c_{buf} / n_{group}$

    $c_{on group}[n_{group}]$
    $D[n_{group}]$
    $D[0] = 0$

    \For{$i \gets 0$ \KwTo $n_{group}$}{
        $c_{on group}[i] = c_{group}$
    }
    
   \For{$i \gets 1$ \KwTo $n_{group}$}{ 
       $D[i] = D[i - 1] + c_{on group}[i - 1]$
   }

    MPI\_Reduce\_scatter from $buf_{send}$ to $buf_{recv}$ at receive offset $r(c_{group})$ of $c_{on group}$ counts on $comm_{group}$\;
    \BlankLine%

    MPI\_Allreduce to $buf_{recv}$ at offset $r(c_{group})$ of $c_{on group}[r]$ counts on $comm_{lane}$
    
   MPI\_Allgatherv to $buf_{recv}$ of $c_{on group}$ counts on $comm_{group}$ using $D$
  
    \caption{\texttt{lane\_allreduce}}\label{alg:lane}
\end{algorithm2e}

Multi-lane collectives~\cite{lane} reduce the variability between message costs through node-awareness.  The multi-lane all-reduce, detailed in Algorithm~\ref{alg:lane}, first performs a reduce-scatter on-node so that each process within the node holds a fraction of the buffer.  Then, each process performs an all-reduce with its corresponding process on each other node.  Finally, the reduced buffer portions are gathered among all processes within a node. 
%Multi-lane versions of collective operations were described in \cite{lane}.  
Each process sending an equal portion of the buffer allows for the operation to better reach injection bandwidth limits since per-process inter-node messages are reduced in both size and count.  Multi-lane algorithms have shown to greatly improve the performance of collectives on older many-core architectures and heterogeneous systems requiring copies of the device buffer to host memory~\cite{gpu_c2c_lane}.  This paper presents a novel extension and analysis of these algorithms to GPU-aware all-reduce operations on heterogeneous systems using GPUDirect RDMA communication; this paper also justifies their continued applicability on modern systems requiring copies between host and device memory. 

\subsection{GPU-Aware All-Reduce Operations}

%Current Ampere and Hopper-based systems have four to eight GPUs per node and up to sixteen CPU cores per GPU. However, existing approaches rely on GPU-to-GPU communication via GPUDirect RDMA or copying data to host memory at each gradient averaging step. While GPUDirect RDMA offloads some of the communication cost to the GPU, it is still highly dependent on CPU and network interface card (NIC) performance. Thus, we propose to share a single GPU across multiple cores of the system, where each core facilitates communication of an equal-sized chunk to its corresponding worker on another GPU. Ideally, this increases the bandwidth utilization of the NICs and Non-uniform memory access (NUMA) region interconnects.
GPU-aware MPI implementations allow the CPU to offload communication to the GPU and network interface card (NIC) by utilizing the GPUDirect RDMA protocol.  The NICs on emerging supercomputers support RDMA, allowing data to flow directly between the NIC and the GPU, bypassing copies with host memory.  %GPUDirect RDMA can perform significantly better than relying on copies in host memory.
Large GPU-aware all-reduce operations take advantage of GPUDirect RDMA.  While manually copying data to the CPU showed benefits on Power9 systems~\cite{perf_hetero}, more modern systems obtain the best performance through the use of GPUDirect RDMA communication. 

Parallel applications that rely on large all-reduces typically utilize a single MPI process per GPU.  To achieve high-performance within the all-reduce, each process performs a GPU-aware all-reduce on the buffer, utilizing GPUDirect RDMA communication at each step of the algorithm.  This requires the entire buffer to be communicated before a kernel is launched to perform the local reduction.  The buffer is then sent on through the next step of the all-reduce only after the entire reduction is complete. 

On many-core CPU systems, similar synchronization overheads have been reduced through partitioned communication, which allows each thread to start communication on a local portion of the buffer without waiting for all threads to finish computation~\cite{partitioned}.  Similarly, partitioned communication allows for portions of the buffer to be received early, so that each thread can begin local computation as soon as its portion of the receive arrives, rather than waiting for the full buffer.  While partitioned communication has been shown to optimize many operations on CPU-only systems, there have been extensions to heterogeneous systems~\cite{gpu_c2c_lane, Temucin_Schonbein_Levy_Sojoodi_Grant_Afsahi_2024}.  Due to the high overheads of kernel launches, along with the inability to initiate communication directly from the GPU, partitioning heterogeneous data into smaller partitions can incur bottlenecks.  

\begin{comment}
\begin{figure*}[ht!]
    \centering
    \begin{subfigure}{0.3\textwidth}
        \centering
        \includegraphics[width=\linewidth,page=1]{Figures/delta_gpu_multi_allgpusactive_pergpu.pdf}
        \caption{(Delta)}\label{fig:delta_microbenchmark}
    \end{subfigure}
    \hfill
    \begin{subfigure}{0.3\textwidth}
        \centering
        \includegraphics[width=\textwidth,page=1]{Figures/tuolumne_gpu_multi_allgpusactive_pergpu.pdf}
        \caption{Default SPX mode (Tuolumne)}\label{fig:spx_microbenchmark}
    \end{subfigure}
    \hfill
    \begin{subfigure}{0.3\textwidth}
        \centering
        \includegraphics[width=\linewidth,page=1]{Figures/tuolumne_cpx_gpu_multi_allgpusactive_pergpu.pdf}
    \caption{Fine-grained CPX mode (Tuolumne)}\label{fig:cpx_microbenchmark}
    \end{subfigure}
    \caption{A ping-pong benchmark using multiple active processes per (logical) GPU featuring performance on Delta (left), and the SPX (middle) and CPX (right) modes of the AMD MI300A on Tuolumne}\label{fig:microbenchmark}
\end{figure*}
\end{comment}

\begin{figure*}[ht!]
    \centering
    \begin{subfigure}{0.32\textwidth}
        \centering
        \includegraphics[width=\linewidth,page=1]{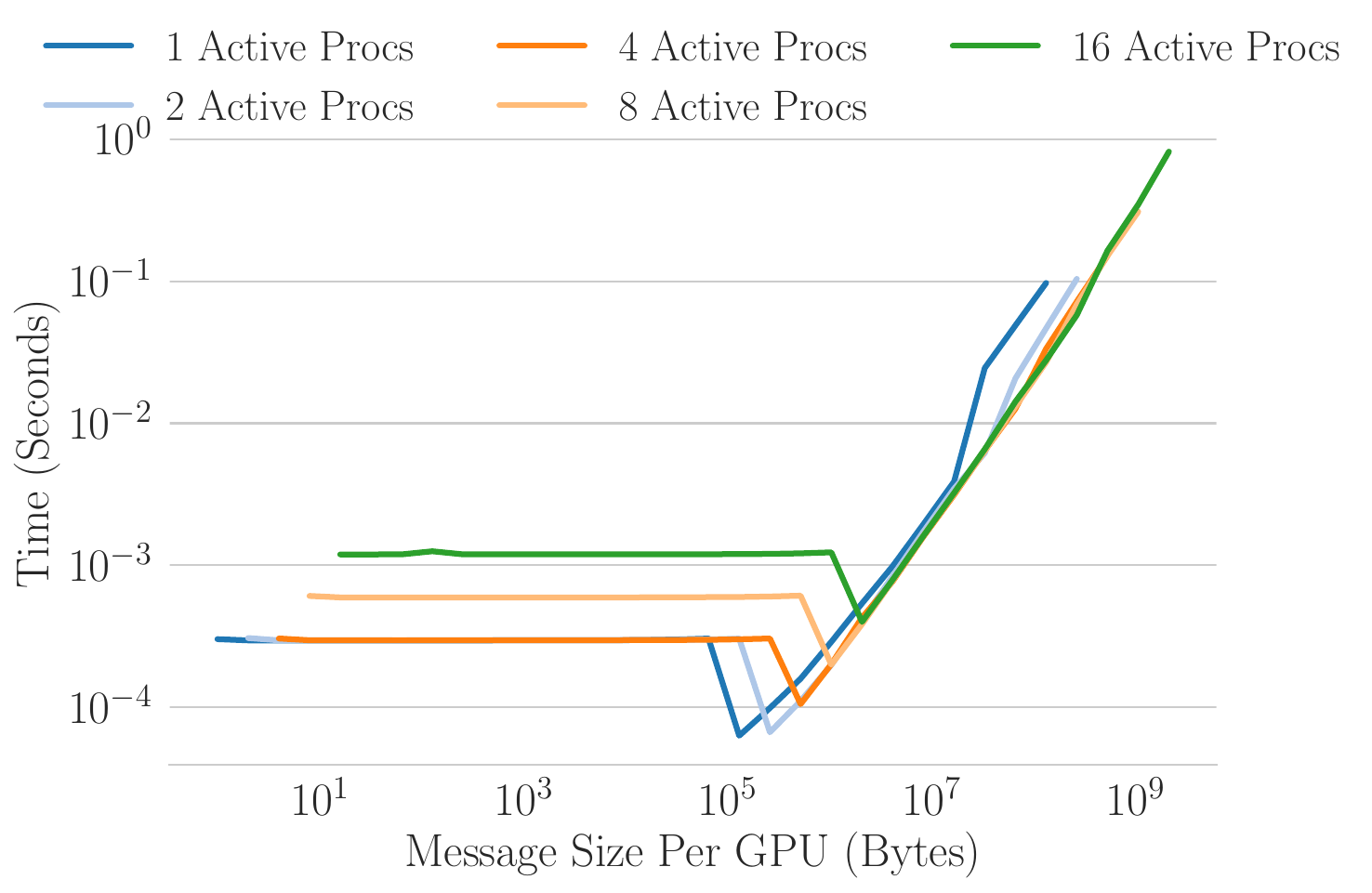}
        \caption{Timings (Delta)}\label{fig:delta_microbenchmark}
    \end{subfigure}
    \hfill
    \begin{subfigure}{0.32\textwidth}
        \centering
        \includegraphics[width=\textwidth,page=1]{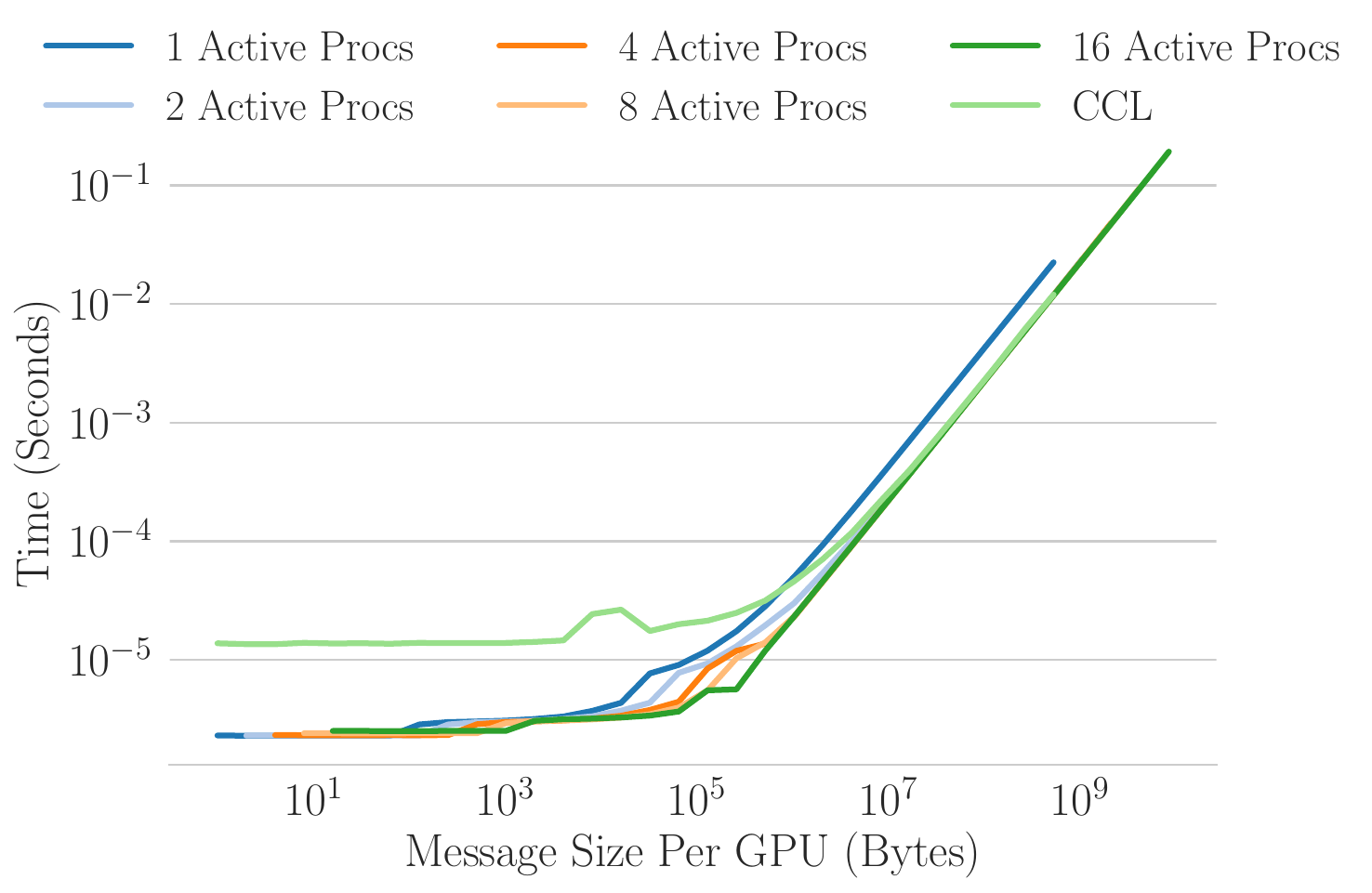}
        \caption{Timings (Tuolumne)}\label{fig:times_microbenchmark}
    \end{subfigure}
    \hfill
    \begin{subfigure}{0.32\textwidth}
        \centering
        \includegraphics[width=\linewidth,page=1]{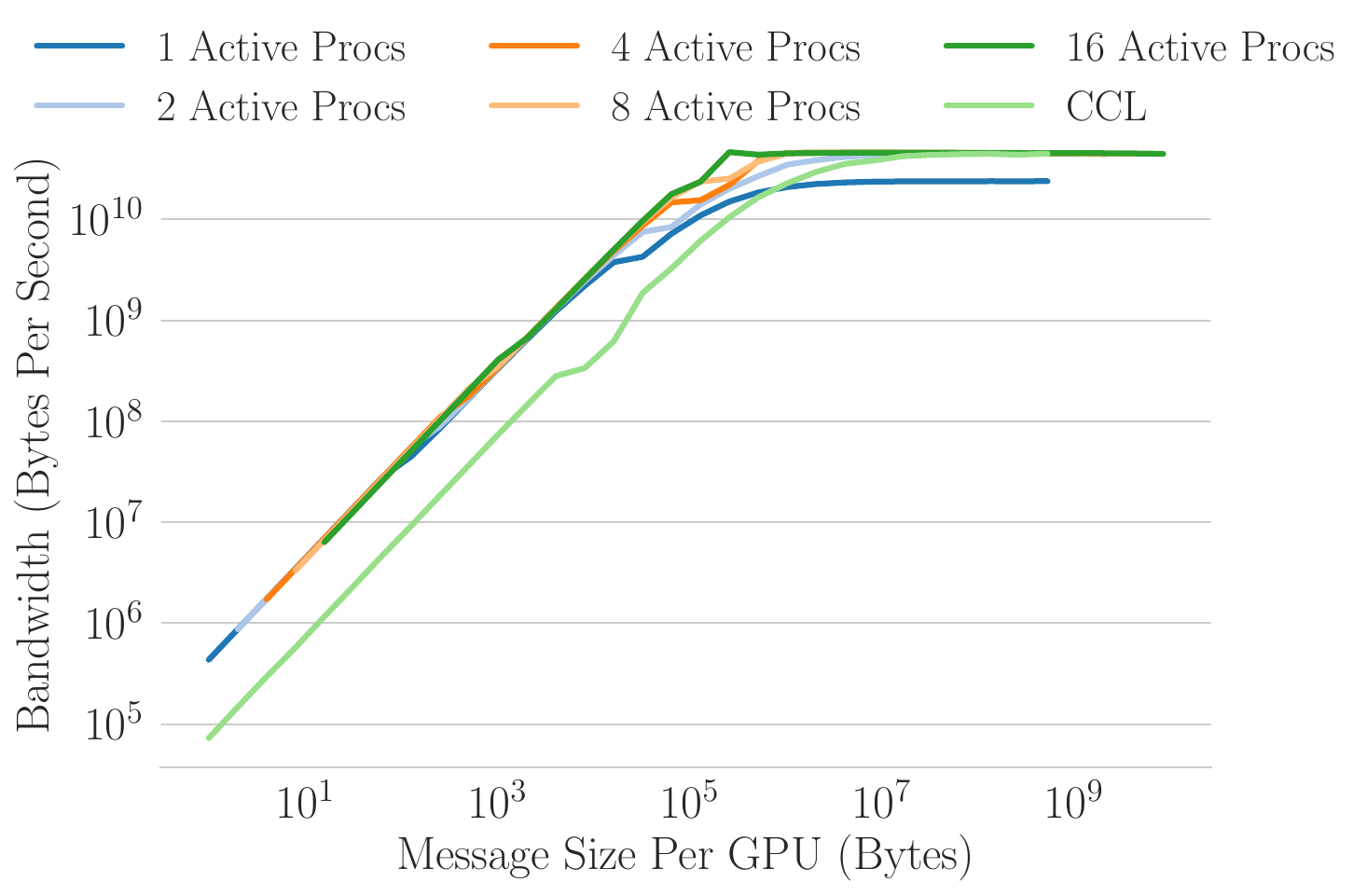}
    \caption{Bandwidth (Tuolumne)}\label{fig:bw_microbenchmark}
    \end{subfigure}
    \caption{A ping-pong benchmark using multiple active processes per (logical) GPU featuring timed performance on Delta (left), and timings (middle) and bandwidth (right) of the AMD MI300A on Tuolumne}\label{fig:microbenchmark}
\end{figure*}

While GPUDirect RDMA offloads some of the communication cost to the GPU, it is still highly dependent on CPU and NIC performance.  Thus, this paper presents a novel optimization for MPI all-reduce operations to share a single GPU across multiple cores of the system, with each advancing a separate portion of the buffer. Ideally, this increases the bandwidth utilization of the NICs and Non-uniform memory access (NUMA) region interconnects while reducing synchronization overheads, allowing for each process to progress as soon as its portion of the buffer is ready.  

% This paper presents a novel optimization for MPI all-reduce operations through the use of multiple processes per GPU, with each advancing a separate portion of the buffer.  This approach reduces synchronizations overheads, allowing for each process to progress as soon as its portion of the buffer is ready.  

\subsection{Partitioning Modes of the AMD MI300A}

The AMD MI300A Accelerated Processing Unit (APU) is a processor that places the CPU and GPU in a single package~\cite{mi300a_overview}. There are three CPU chiplets each with 8-core AMD Zen4 CPUs and six Accelerator Complex Dies (XCDs) with 38 CDNA3-based GPU compute units each.  128 GB of HBM3 memory is accessible to both CPU cores and GPU compute units backed by 256 MB of shared last level cache.

At runtime, the XCDs can be grouped logically depending on the user's needs. By default, the MI300A APU operates in Single Partition X-celerator (SPX) mode in which all 6 XCDs are exposed to the user as a single GPU. In this case, the workload on the GPU is automatically distributed across all XCDs.  With the Triple Partition X-celerator (TPX) mode, the APU is treated as three GPUs of two XCDs each. In Core Partitioned X-celerator (CPX) mode, each XCD is exposed as a single GPU.

The TPX and CPX partitioning modes offer the developer more granular control over the hardware that is chosen to execute their applications.  Furthermore, since all partitioning modes can access the entire pool of HBM3 memory, these modes offer increased parallelism when used effectively.

\subsection{Related Work}
\begin{comment}
Here are some ideas, I can help you find citations of these and can also point you to additional related works but this should help you get started:
\begin{itemize}
    \item Related collectives 
    \item Node-aware collective optimizations (I have a node-aware allreduce paper for small allreduce operations)
    \item Topology-aware collectives
    \item Locality-aware aggregation in other contexts (I/O, neighborhood collectives, p2p communication)
\end{itemize}
\end{comment}

% \todo[inline]{locality-aware collectives}

Other collectives such as the all-to-all, all-gather and related neighborhood methods feature a similar communication pattern to the all-reduce where in the worst case, every pair of processes must communicate. Intra-node communication has been optimized with shared memory~\cite{shared_memory,GrahamSharedMemoryColl2008,JainSharedMemoryColl2018} on CPUs. Hierarchical collectives aggregate all data onto a single leader per node, reducing inter-node traffic~\cite{hierarchical, KaronisHierColl2000,GrahamHierCheetah2011,TraffHierAllgather2006}.  Multi-leader collectives perform hierarchical algorithms with multiple leaders per node~\cite{multileader_hierarchical,multileader_hierarchical2}.  Locality-aware aggregation has also been used to aggregate messages, reducing the number of inter-node steps, both in the context of small all-reduce operations~\cite{bienz_allreduce}, as well as other collective operations~\cite{bienz_bruck_allgather, locality_aware_alltoall}.

Topology-aware approaches for collective operations have also been developed. For instance, for a cluster layout consisting of a two-dimensional grid of nodes where each node is directly connected to its four neighboring nodes, an all-reduce can be performed twice, once in each axis direction~\cite{BhateleTorusColl2012, torus}, approximately reducing the number of communicating nodes from the total number of nodes to the sum of all axes dimensions with the cost of an additional all-reduce.  Topology-aware mapping is commonly used to map processes within other topologies~\cite{topo0, topo1,topo2}, such as trees~\cite{PatarasukTreeAllreduce2007}, to minimize long-distance messages.

GPU-aware optimizations for collectives have been widely explored in recent years.  Hierarchical methods have been analyzed on heterogeneous architectures~\cite{infini_hierarchical, gpu_multileader_hierarchical, hierarchical_ccl}, communication primitives and collectives have been optimized to use direct memory access between GPUs more efficiently~\cite{gpu_dma, gpu_dma2,gpu_dma3,gpu_dma4}, and compressed collectives have been used to reduce communication costs~\cite{comp_cosmo,comp_alltoall,comp_allreduce,zccl}.

Proprietary communication libraries, such as NVIDIA's Collective Communications Library (NCCL)~\cite{NVIDIA2025NCCL} and AMD's ROCm Communication Collectives Library (RCCL)~\cite{AMD2025RCCL}, implement the all-reduce with increased knowledge about the GPU and NIC hardware in use, yielding optimizations over MPI implementations that make them commonplace in artificial intelligence (AI) applications.  These all-reduce operations achieve very high bandwidth, optimal for large reductions.  Similar to MPI, these all-reduces are typically performed with a single process per GPU. 

\section{Microbenchmarks}\label{sec:microbenchmarks}

GPU-aware MPI operations, such as the all-reduce, are performed using one controlling process per GPU. This process initiates all GPU-aware communication involving its GPU. However, modern systems feature many more CPU cores than GPUs, causing them to be underutilized. 

To demonstrate the potential advantage of using multiple processes per GPU, we evaluated a ping-pong benchmark on both systems analyzed. In this benchmark, there are two nodes where each rank on a node sends and receives a message with its corresponding rank on the other node. Each GPU on a node is participating in the benchmark and controlled by at least 1 MPI process. To obtain the time it takes to send or receive one message, the round-trip time is divided by $2$. 

\begin{comment}
\begin{figure}[ht!]
    \centering
    \includegraphics[width=\linewidth,page=1]{Figures/delta_gpu_multi_allgpusactive_pergpu.pdf}
    \caption{A ping-pong benchmark on Delta using multiple processes per GPU}\label{fig:delta_microbenchmark}
\end{figure}
\end{comment}

Benchmark results on the Delta system with intermediate copies between host and device memory can be seen in Figure~\ref{fig:delta_microbenchmark}. When 16 processes per GPU are used, this yields a significant speedup of $3.49$x over a standard ping-pong of the largest message size, suggesting a benefit by the use of multiple processes per GPU during collective communication on Delta.

Results using GPUDirect RDMA communication on Tuolumne with the APU in SPX mode can be seen in Figure~\ref{fig:times_microbenchmark} with timings of the communication and Figure~\ref{fig:bw_microbenchmark} with the communication bandwidth.  Similar to results on Delta, the ping-pong yields speedups of up to $1.89$x over a standard ping-pong when using 16 processes per GPU. 
%Therefore, these results suggest using mutiple processes per GPU using MPI is applicable on Tuolumne's APUs.
Further, MPI message passing routines are unable to fully saturate the available bandwidth with a single process per GPU, as shown in Figure~\ref{fig:bw_microbenchmark}.  However, when multiple processes communicate data from a single GPU, these maximum bandwidth limits are achieved.  While RCCL has lower bandwidth than MPI for smaller messages, it is able to fully saturate available bandwidth for large exchanges with only a single process per GPU. 
%As illustrated in both plots, RCCL already achieves the best performance at the largest message sizes by reaching injection bandwidth.

\section{Methods}\label{sec:methods}
\begin{comment}
\begin{itemize}
    \item Main idea of novel algorithms: each process wants to communicate only with corresponding process on every other node (e.g. process controlling GPU0 on node n communciates only with processes controlling GPU0 on other nodes).
    \item May want to put the approaches in an algorithm environment.  
    \item Approach 1: reduce-scatter + allreduce + allgather
    \item Approach 2: allreduce on node + allreduce off node
    \item Analysis of these approaches.  Reduce scatter minimizes size of off node communication, each process per node performs a reduce on 1/gpus\_per\_node (or 1/procs\_per\_node) portion of data.  However this approach requires an additional allgather at the end, increasing the amount of on node communciation.
    \item Approach two reduces the number of steps of the allreduce.  Original approach, each process sends data in a ring p-1 times.  New approach: only (ppn-1 + nnodes-1) steps of communication.
\end{itemize}
\end{comment}

Baseline GPU-aware all-reduce operations implement standard algorithms, such as the ring and multi-lane implementations in Algorithms~\ref{alg:ring} and~\ref{alg:lane}.  These baseline implementations use a single MPI process per GPU with a buffer size $s$.  

\subsection{Multiple Processes Per GPU}
We optimize the baseline collective operations by using multiple processes per GPU with the all-reduce serving as a proof-of-concept. Each GPU on a node is assigned to an equal number of unique processes, as exemplified in Figure~\ref{fig:allreduce_ppg}.  In this example, there are two processes per GPU, displayed as red and blue outlines. The buffers to be reduced are partitioned evenly across the multiple processes per GPU.  
%It is assumed that each node has the same number of cores and GPUs such that the distinct sections can be reduced between nodes using all cores.  
%The consistent number of processes also allows the on node stages between GPUs to utilize all cores, where applicable.

Assuming there are \texttt{PPG} processes per GPU, each process $r$ on a node has a local rank $l_{r} \leftarrow r \mod{\texttt{PPG}}$.  Each process with local rank $l_{r} = 0$ is assigned as a leader process.  The leader process creates the full buffer to be reduced on its corresponding device.  After the buffer is created, the leader extracts its Inter-Process Communication (IPC) memory handle with \texttt{(cuda/hip)GetIPCMemHandle}.  This memory handle is then broadcast to all processes assigned to the given GPU.  Each non-leader process gains access to the shared device pointer by opening the received memory handle, using \texttt{(cuda/hip)IpcOpenMemHandle}.

Before an all-reduce operation can occur, new communicators must be created. A communicator, \texttt{new\_comm}, must be created for all processes with equal local rank $l_{r}$, allowing for all-reduce operations on each communicator to reduce equivalent fractions of the buffer among all GPUs.  All processes with local rank $l_{r}$ operate on the same fraction of a buffer located at $buf[\frac{s}{\texttt{PPG}}*l_r]$, but are assigned to separate GPUs.  For example, all processes outlined as red in Figure~\ref{fig:allreduce_ppg} will reduce the first half of all buffers, while the blue processes reduce the second half. 

\begin{comment}
\begin{figure*}[ht!]
    \centering
    \begin{subfigure}{0.49\textwidth}
        \centering
        \includegraphics[width=\textwidth,page=13]{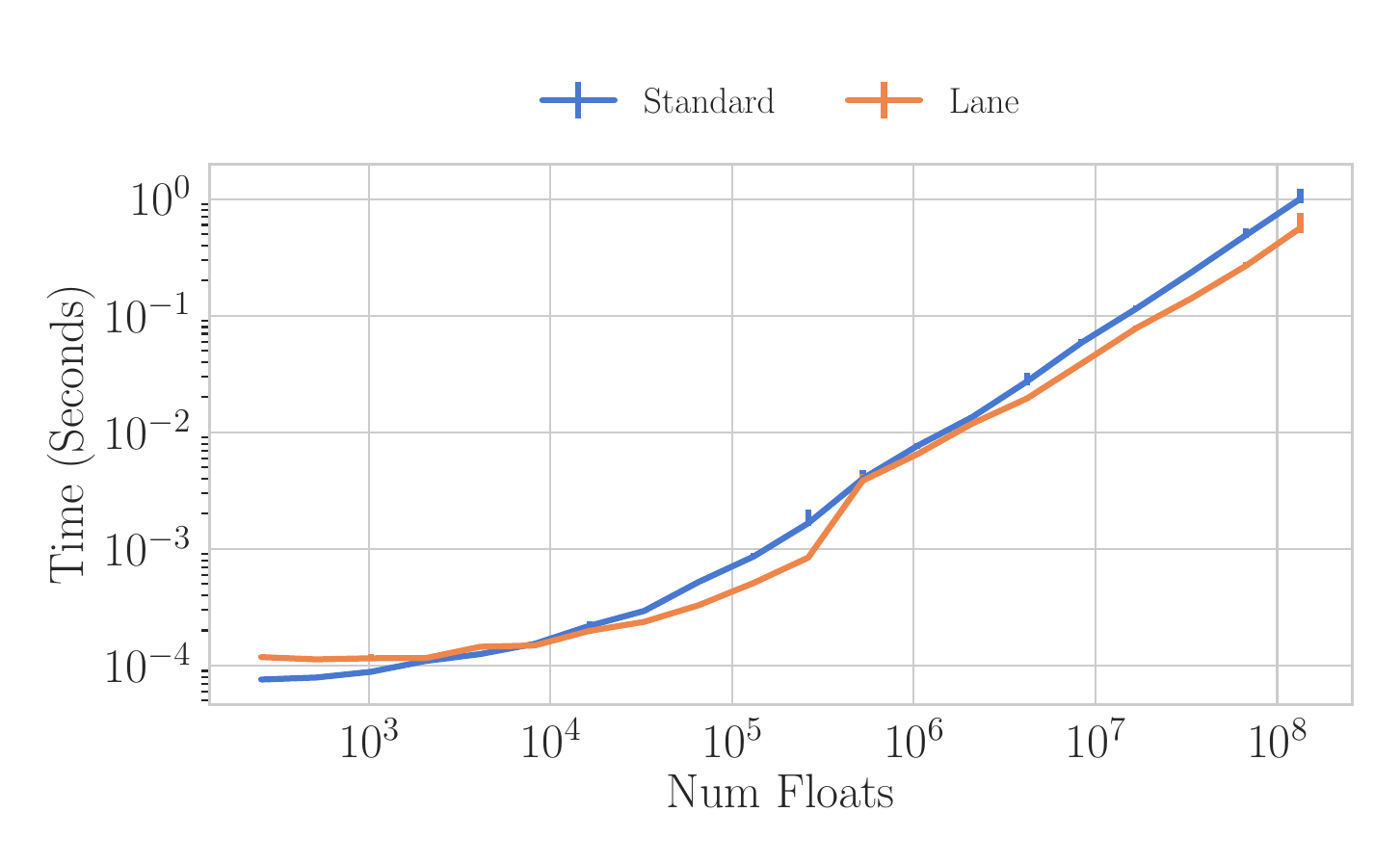}
        \caption{8 nodes, varying buffer sizes}
    \end{subfigure}
    \hfill
    \begin{subfigure}{0.49\textwidth}
        \centering
        \includegraphics[width=\textwidth,page=16]{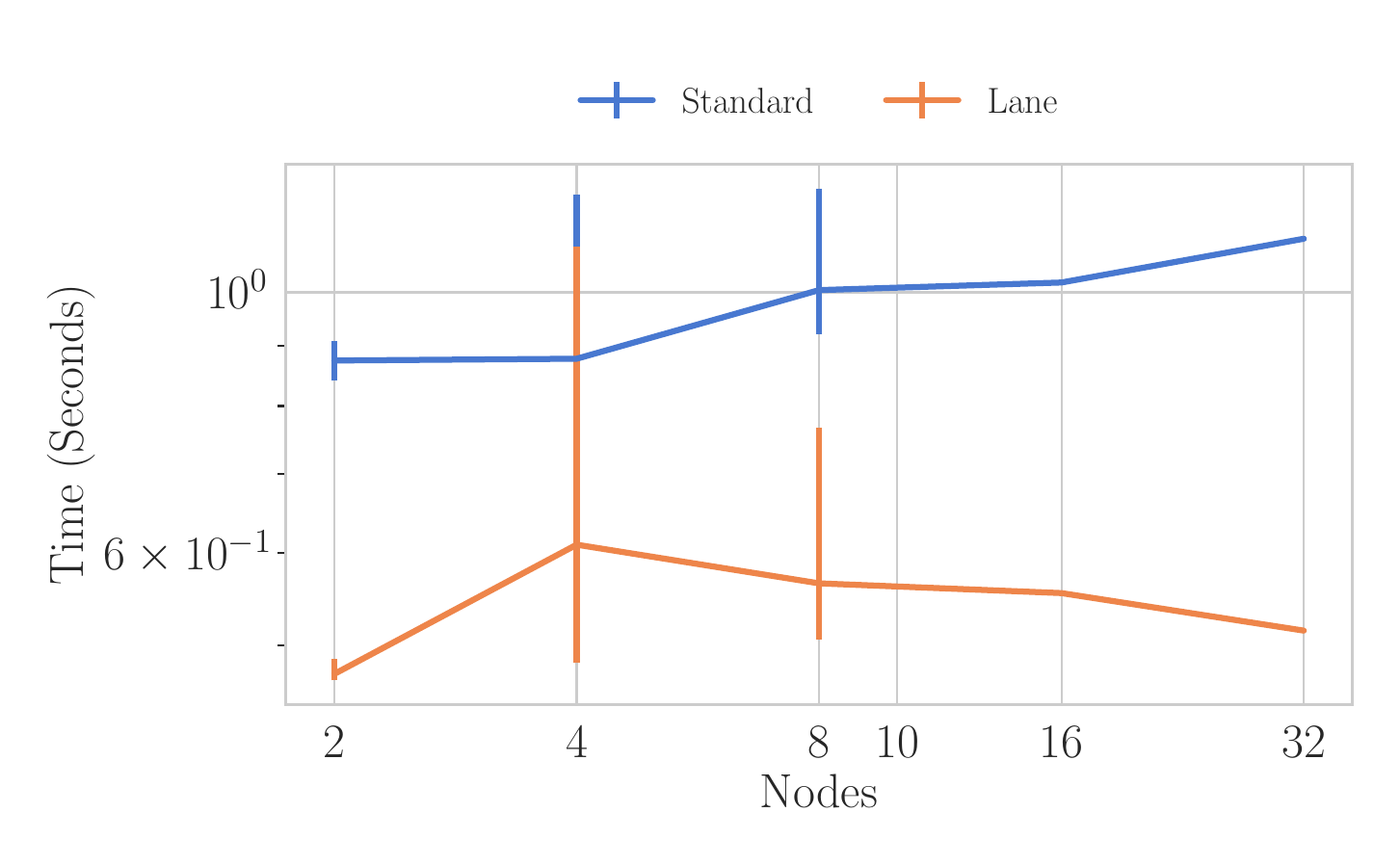}
        \caption{$2^{27}$ floats, varying node counts}
    \end{subfigure}
    \caption{Multi-Lane GPU-Aware Performance on Delta}\label{fig:std_vs_lane}
\end{figure*}
\end{comment}
\subsubsection{Standard Approach}
\begin{figure}[ht!]
    \centering
    \includegraphics[width=\linewidth]{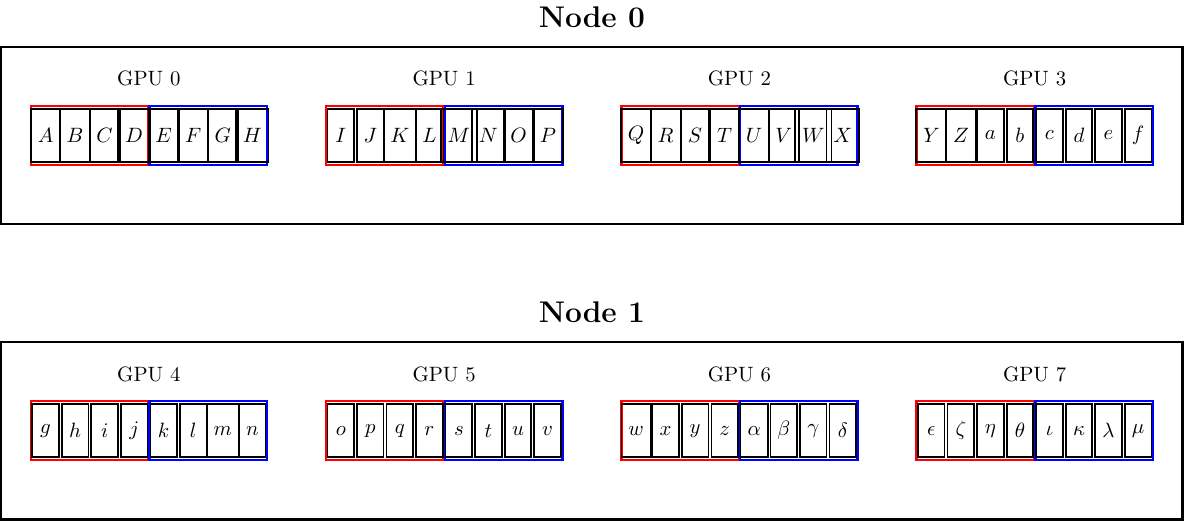}
    \caption{Partitioning Buffer Over Multiple Processes Per GPU}
    %\Description{}
    \label{fig:allreduce_ppg}
\end{figure}

After the initial setup, each process performs an equal portion of the all-reduce asynchronously.  
%Each process per GPU is assigned a contiguous chunk of both the send and receive buffer. 
Processes assigned to the same offset in the buffer of each GPU perform an all-reduce.  For example, two all-reduces are performed concurrently in Figure~\ref{fig:allreduce_ppg}; one among all red processes, and a second among all blue processes.

In the baseline case with only a single process per GPU, all processes would reduce a buffer of $s$ bytes with the following.
\begin{lstlisting}
MPI_Allreduce(sendbuf, recvbuf, s, MPI_SUM,
    MPI_COMM_WORLD);
\end{lstlisting}
When using multiple processes per GPU, this all-reduce call becomes the following.
\begin{lstlisting}
MPI_Allreduce(&(sendbuf[(s/PPG)*l_r]), 
    &(recvbuf[(s/PPG)*l_r]), 
    s/PPG, MPI_SUM, new_comm);
\end{lstlisting}

%The collective communication primitive, the Allreduce, is used in large data parallel deep learning applications. Workers, or GPUs in the case of deep learning, perform a reduction operation (summation) over a set of shared floating point numbers and then share the final result to each other.

%First, we describe an algorithm where every worker communicates with another corresponding worker on all other GPUs, the standard approach. Next, we divide the communicated data by the number of processes per node with an on node Reduce-scatter and Allgatherv between corresponding processes of all on node GPUs, the lane-awarw approach.

%For each approach, we setup our experiments as follows.  

In this algorithm, the total number of messages among all processes increases by the factor of the number of processes per GPU. However, each message is reduced in size by the same factor. These messages are communicated concurrently.

The number of kernel launches per process remains constant.  Data transfers using host copy routines are performed by the CPU core that each MPI rank was bound to at process launch, likely leading to parallelization. The Multi-Process Service (MPS) is disabled on the Ampere-based system, so copies between host and device memory are serialized.  
%For GPUDirect RDMA routines, communication is parallelized using MPS when multiple processes compete for the GPU at once.  However, this contention is unlikely due to the natural asynchrony among processes.
%For GPUDirect RDMA routines, communication is either parallelized using MPS or, as there is asynchrony among processes, all kernels are unlikely to compete for the GPU at one time.  
On the AMD system tested, GPUDirect RDMA routines are used. MPS is implicitly enabled; equivalent features are natively supported in the GPU driver~\cite{AMDMPSON}, parallelizing GPU tasks dynamically at runtime. 
%In our tests, communication is driven by either (a) the MPS-enabled GPU driver parallelizing tasks dynamically at runtime when GPUDirect RDMA is used, or (b) the utilization of the CPU core bound to each MPI rank at process launch when host copy routines are used.
%Since the GPU kernel calls of all processes assigned to a GPU are approximately completed in parallel, the kernel overhead is not increased significantly. 
%The NVIDIA tool, Inter-Process Communication (IPC), handles the on GPU concurrency such that, after buffer broadcasting, process synchronization is simple. 
%With GPUDirect RDMA, communication is driven by the MPS-enabled GPU driver parallelizing tasks dynamically at runtime whereas host copy routines use the CPU core that was bound to each MPI process at process launch to drive communication. 

\subsubsection{Multi-lane Approach}
The standard approach for using multiple processes per GPU can be further extended to utilize multi-lane collectives. In this example, the multi-PPG all-reduce is broken into three stages: 1) intra-node reduce-scatter, 2) inter-node all-reduce on a lane communicator, and 3) intra-node all-gatherv, as described in Algorithm~\ref{alg:lane}.  In the multi-process version of the multi-lane algorithm, the send buffer, $buf_{send}$, and receive buffer, $buf_{recv}$, are passed in at offset $\frac{s}{\texttt{PPG}}*l_r$.  Further, $comm_{group}$ contains corresponding processes of on-node GPUs and $comm_{lane}$ contains corresponding processes on other nodes.  To create these communicators, \texttt{new\_comm} is split into processes that reside on the same node ($comm_{group}$) and processes that reside on separate nodes ($comm_{lane}$).   With multiple processes per GPU, each CPU core is a separate lane. 
%The reduce-scatter and all-gatherv stages occur between the respective cores per GPU on node.  

\begin{figure}[ht!]
    \centering
    \includegraphics[width=\linewidth]{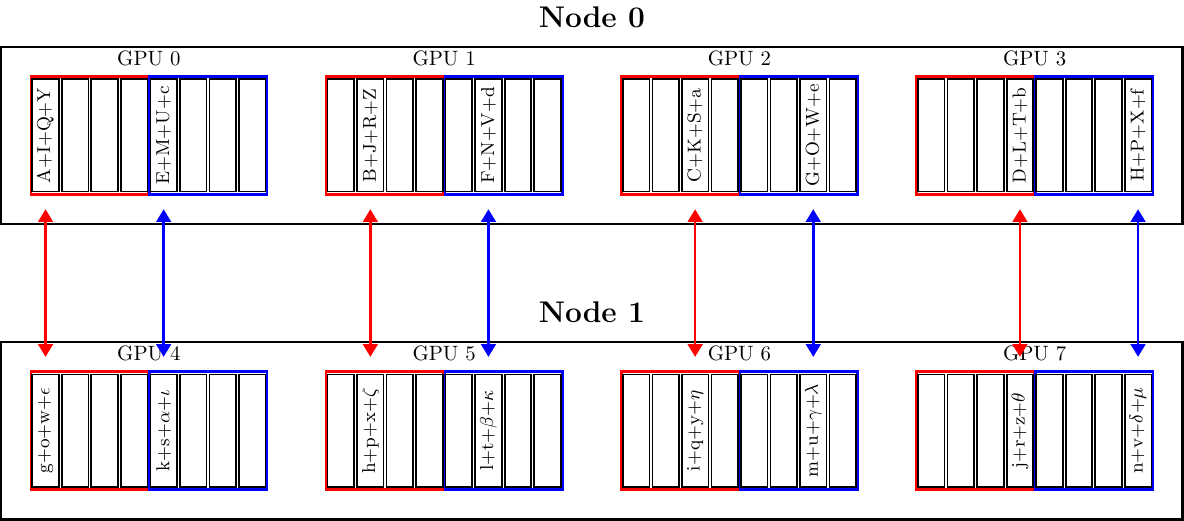}
    \caption{Multi-\texttt{PPG}, Multi-Lane}
    %\Description{}
    \label{fig:allreduce_ppg_lane}
\end{figure}

The inter-node stage of this all-reduce is displayed in Figure~\ref{fig:allreduce_ppg_lane}.
Initially, all red processes perform intra-node reduce-scatters.  At the same time, blue processes perform separate intra-node reduce-scatters.  After the scattered buffers are reduced, each process performs inter-node all-reduce operations with corresponding processes.  Finally, red and blue processes concurrently gather their reduced data on-node.

%$\left(r_n \mod ppg\right)*\left(\frac{count}{ppg}\right)$ where $r_n$ is the on node rank and $ppg$ is the number of processes per GPU.

The reduce-scatter divides the input buffer into chunks equal to the number of GPUs on a node.  The off-node all-reduce messages are reduced in size proportionally, leading to a total reduction in inter-node buffer size by the number of MPI processes per node. However, there is an increase in intra-node communication from the reduce-scatter and the subsequent all-gatherv operations. 

\section{Results}\label{sec:results}
\begin{comment}
\begin{itemize}
    \item Setup problem: each computer that was tested, details about the computer (e.g. Grace Hopper, 4 NVIDIA xxx GPUs per node, Intel MPI).
    \item General tests: time to perform an allreduce of various sizes across all GPUs.  
    \item Include results for a few computers here, both for GPU-aware MPI and NCCL
    \item Analysis of results
    \item May be interesting to time each of the portions of each allreduce separately (e.g. Approach 1 time reduce-scatter vs allreduce vs allgather at large scale... where is the time spent?  Same for approach 2, time each allreduce independently to see breakdown of cost).  This may imply that the bottleneck is in the allgather or something similar, and a better implementaiton may lead to further improvements.
\end{itemize}
\end{comment}

% We can list the details of our runs in this section
%We perform performance analysis of our Allreduce algorithms on two systems:  Delta at UIUC with 4 NVIDIA A100s and 64 cores per node, as well as Tioga and Tuolumne at LLNL. \todo[inline]{Maybe list specs of LLNL systems} \todo[inline]{TODO: Are we testing Bridges-2 V100's too?} 

%We show that using all available cores per GPU on large buffers yields an X to Y times speedup on X and Y nodes of Delta using OpenMPI, respectively. On Tioga using RCCL, we see...
\enlargethispage{1.3\baselineskip}
The impact of multi-lane and multi-process per GPU all-reduce algorithms is analyzed against OpenMPI and Cray MPICH on state-of-the-art supercomputers: NCSA's Delta and LLNL's Tuolumne.  Some details of these systems are provided below.

\textbf{Delta: } This system provides a variety of nodes, including both NVIDIA and AMD GPUs.  The number of AMD nodes is limited, so only NVIDIA nodes were benchmarked for this paper.  All-reduce operations were analyzed on the single-socket nodes, each composed of a 64 core AMD Milan CPU with 4 NVIDIA A100 GPUs.  

\textbf{Tuolumne: } Each node has four AMD MI300A APUs, each of which has a 24-core AMD EPYC CPU and a CDNA 3 GPU.

The impact on MPI all-reduce operations using copies between host and device memory is analyzed solely on Delta, as only OpenMPI was available on this system at the time of writing. Tuolumne supports GPUDirect RDMA provided by Cray MPICH. Cray MPICH does not support any buffer from being communicated through GPU-aware MPI operations if it was not allocated by the calling process.  As a result, sharing IPC buffers within GPU-aware MPI is not possible on these systems.  Thus, each reduction using multiple processes per GPU analyzed on Tuolumne copies to and from another device buffer allocated by the calling process before calling the MPI routines on the local buffer.

%We performed benchmarks of reducing buffers of up to $2^{27}$ 32-bit floating point numbers on   The parallel systems are detailed as follows.

%  NCCL 2.19.3 and CUDA 11.8.0 were also used for testing the standard approach. \todo[inline]{NCCL: List scales/\# of tests}

%Two Allreduce algorithms are used in the off node reduction phase of the multi-core-aware GPU-lane Allreduce. First, the reference MPI\_Allreduce implementation for a given supercomputing platform serves as a baseline. The Baidu ring Allreduce algorithm \cite{ring}, seen in Algorithm X, assumes a ring topology such that communication by each rank is performed only with its neighbors.

Note that NVIDIA MPS was not enabled during benchmarking on Delta whereas an equivalent implementation by AMD is implicitly enabled on Tuolumne.

\begin{figure}[ht!]
    \centering
    \includegraphics[width=\linewidth,page=13]{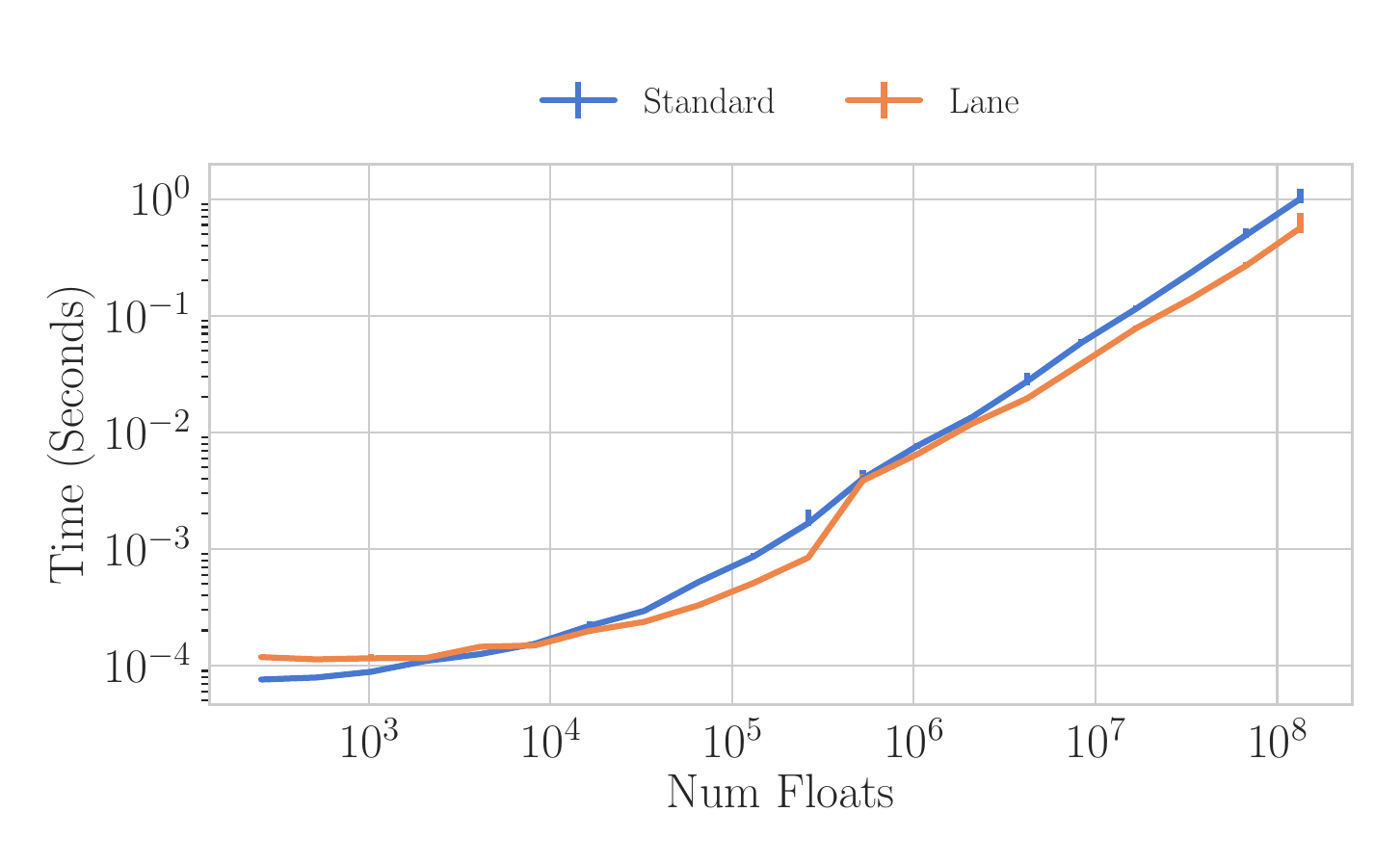}
    \caption{Multi-lane GPU-aware performance on 8 nodes of Delta, varying buffer sizes}\label{fig:std_vs_lane1}
\end{figure}

%\todo[inline]{List Tioga and Tuolumne specs/\# procs/scales/\# of tests. Mention why MPI is not tested on the other machines - Cray MPICH does not support CUDA and ROCm IPC buffers.}  When benchmarking RCCL on xx, a few tests were performed with the environment variables \texttt{NCCL\_\-MIN\_NCHANNELS} and \texttt{NCCL\_\-MAX\_NCHANNELS} set to 4 in order to eliminate the case of low GPU core utilization during memory operations; no differences were observed in timings.

\subsection{Host Copy Analysis}\label{sec:host_copy_results}
All all-reduce tests on Delta were analyzed using OpenMPI 5.0.5 and CUDA 12.4.0 using copies to and from host memory managed by OpenMPI.  All processes were assigned to the GPU adjacent to their NUMA region.  Each benchmark was run $10$ separate times to demonstrate reproducibility.  All plots contain error bars showing the run to run variation of the algorithms.

\begin{figure}[ht!]
    \centering
    \includegraphics[width=\linewidth,page=16]{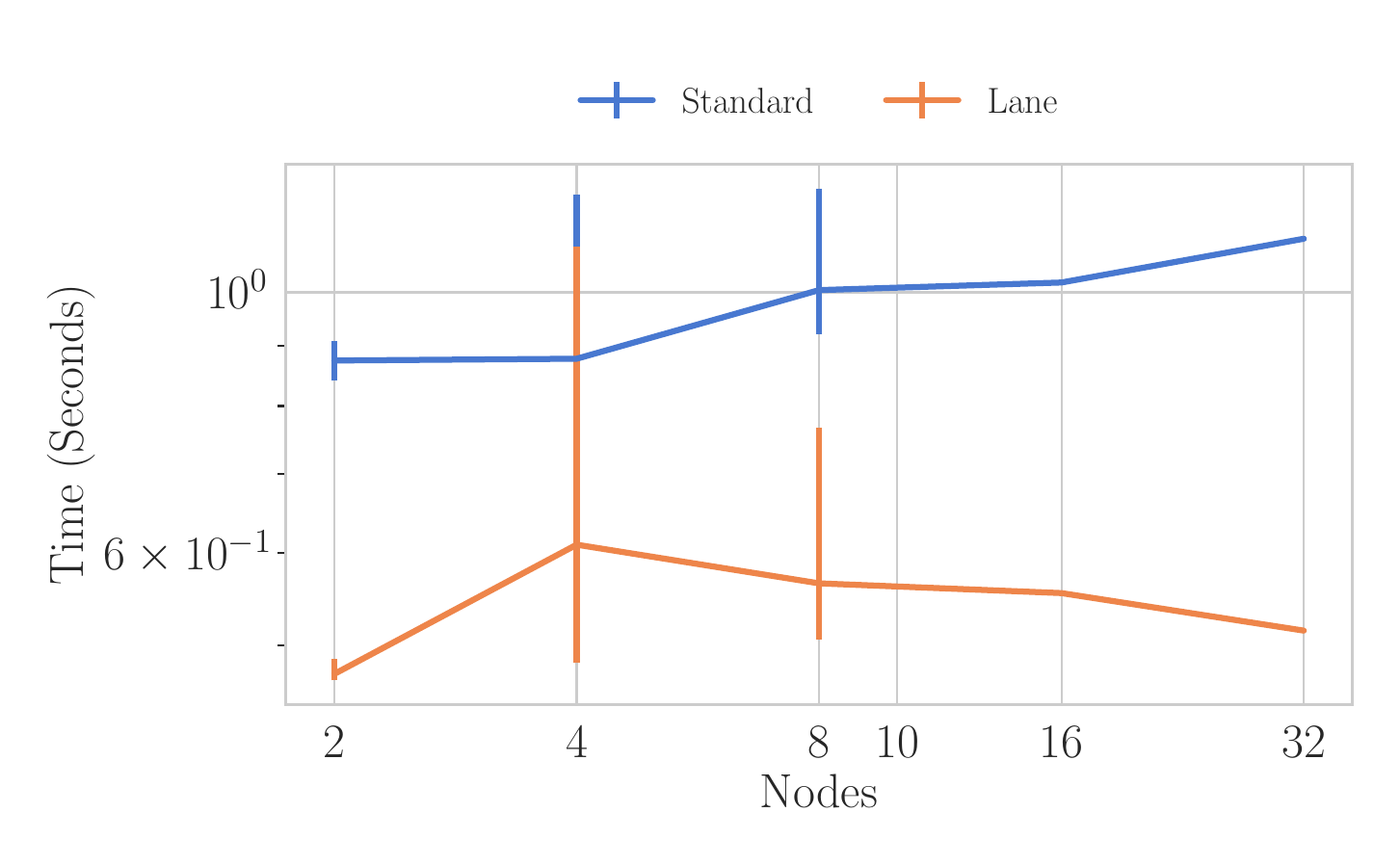}
    \caption{Multi-lane GPU-aware performance of 134 million floats on Delta, varying node counts}\label{fig:std_vs_lane2}
\end{figure}
Figures~\ref{fig:std_vs_lane1} and \ref{fig:std_vs_lane2} compare the performance of a standard \texttt{MPI\_\-Allreduce} operation against the multi-lane optimization.  These tests are using a single process per GPU, and using all four GPUs available per node on Delta.  While the standard algorithm outperforms multi-lane at the smallest data sizes, the multi-lane algorithm yields significant speedups for larger all-reduces.  The speedups for the multi-lane algorithm increase with node count, achieving $2.1$x speedups at $32$ nodes.  

\begin{comment}
\begin{figure*}[ht!]
    \centering
    \begin{subfigure}{0.49\textwidth}
        \centering
        \includegraphics[width=\textwidth,page=11]{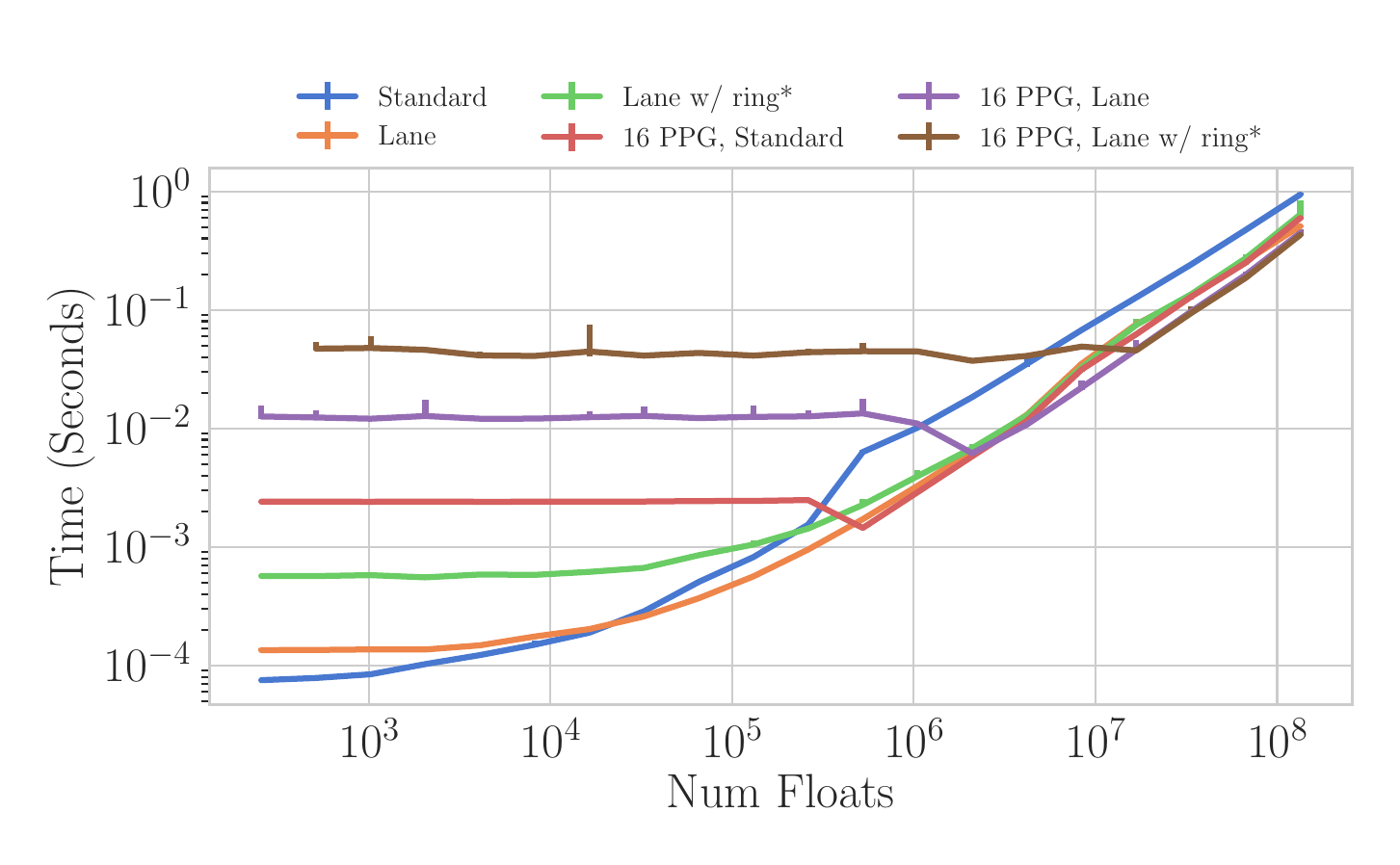}
        \caption{8 nodes, varying buffer sizes}
    \end{subfigure}
    \hfill
    \begin{subfigure}{0.49\textwidth}
        \centering
        \includegraphics[width=\linewidth,page=11]{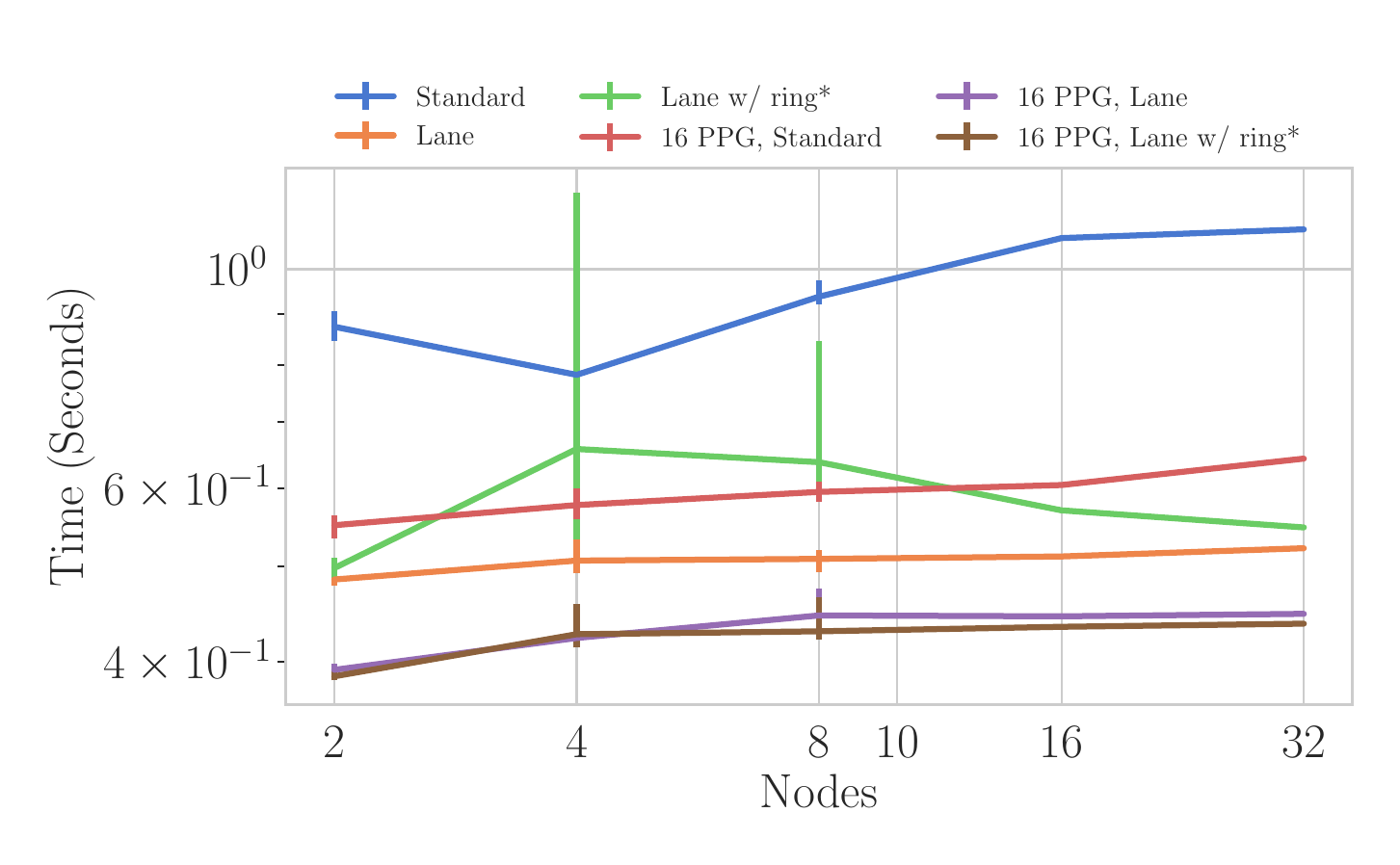}
    \caption{$2^{27}$ floats, varying node counts}
    \end{subfigure}
    \caption{Standard and multi-lane (with \texttt{MPI\_Allreduce} and ring all-reduce) algorithms with 16 processes per GPU on Delta}\label{fig:full_mpi_comparison}
\end{figure*}
\end{comment}

\begin{figure}[ht!]
    \centering
    \includegraphics[width=\linewidth,page=9]{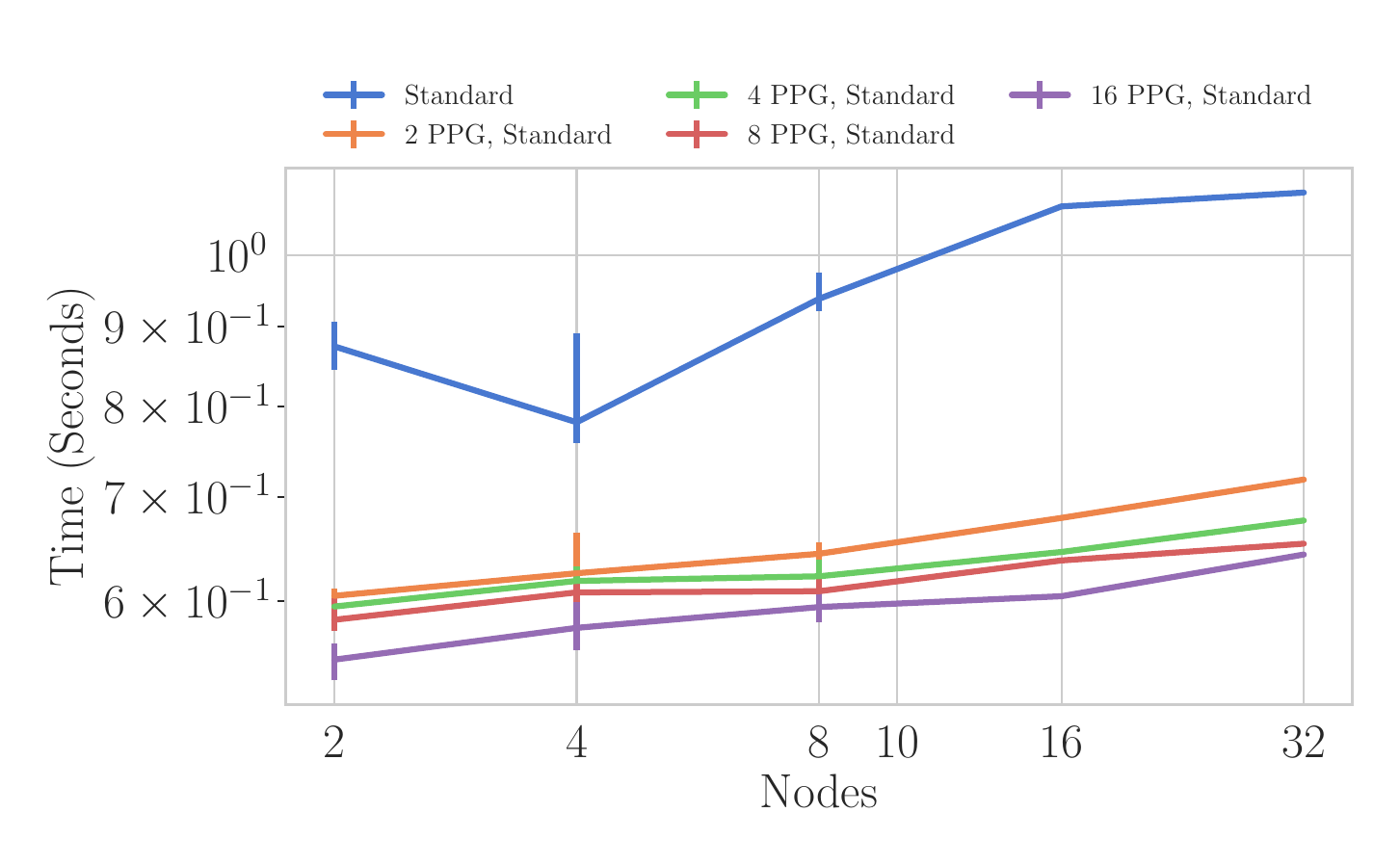}
    \caption{Standard algorithm with multiple processes per GPU of 134 million floats on Delta, varying node counts}\label{fig:mult_ppg}
\end{figure}

\begin{figure}[ht!]
    \centering
    \includegraphics[width=\linewidth,page=5]{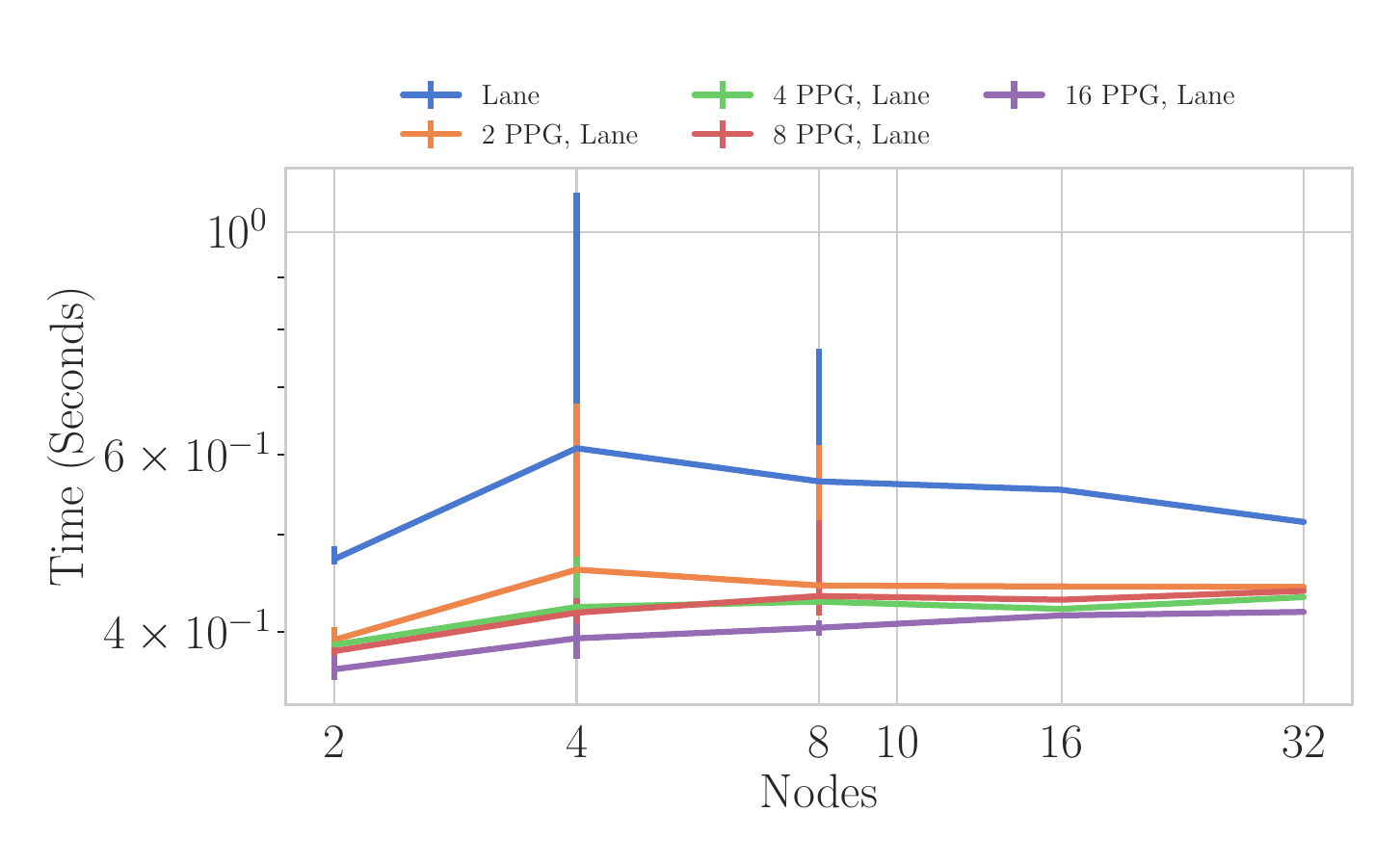}
    \caption{Multi-lane algorithm with multiple processes per GPU of 134 million floats on Delta, varying node counts}\label{fig:mult_ppg_lane}
\end{figure}

Figure~\ref{fig:mult_ppg} shows the performance of a standard \texttt{MPI\_Allreduce} when varying the number of processes per GPU from $1$ to $16$.  Performance continually improves as the number of processes per GPU increases.  Further, these speedups increase with node count through $16$ nodes, yielding over $1.7$x speedup at $32$ nodes.

Figure~\ref{fig:mult_ppg_lane} displays the performance of the multi-lane algorithm when using multiple processes per GPU.  Similar to the standard \texttt{MPI\_Allreduce}, the multi-lane algorithm is further accelerated with increasing numbers of processes per GPU.  Large all-reduce operations across all $4$ GPUs on $8$ nodes yield around a $1.39$x speedup over the multi-lane algorithm with a single process per GPU.

\begin{figure}[ht!]
    \centering
    \includegraphics[width=\linewidth,page=11]{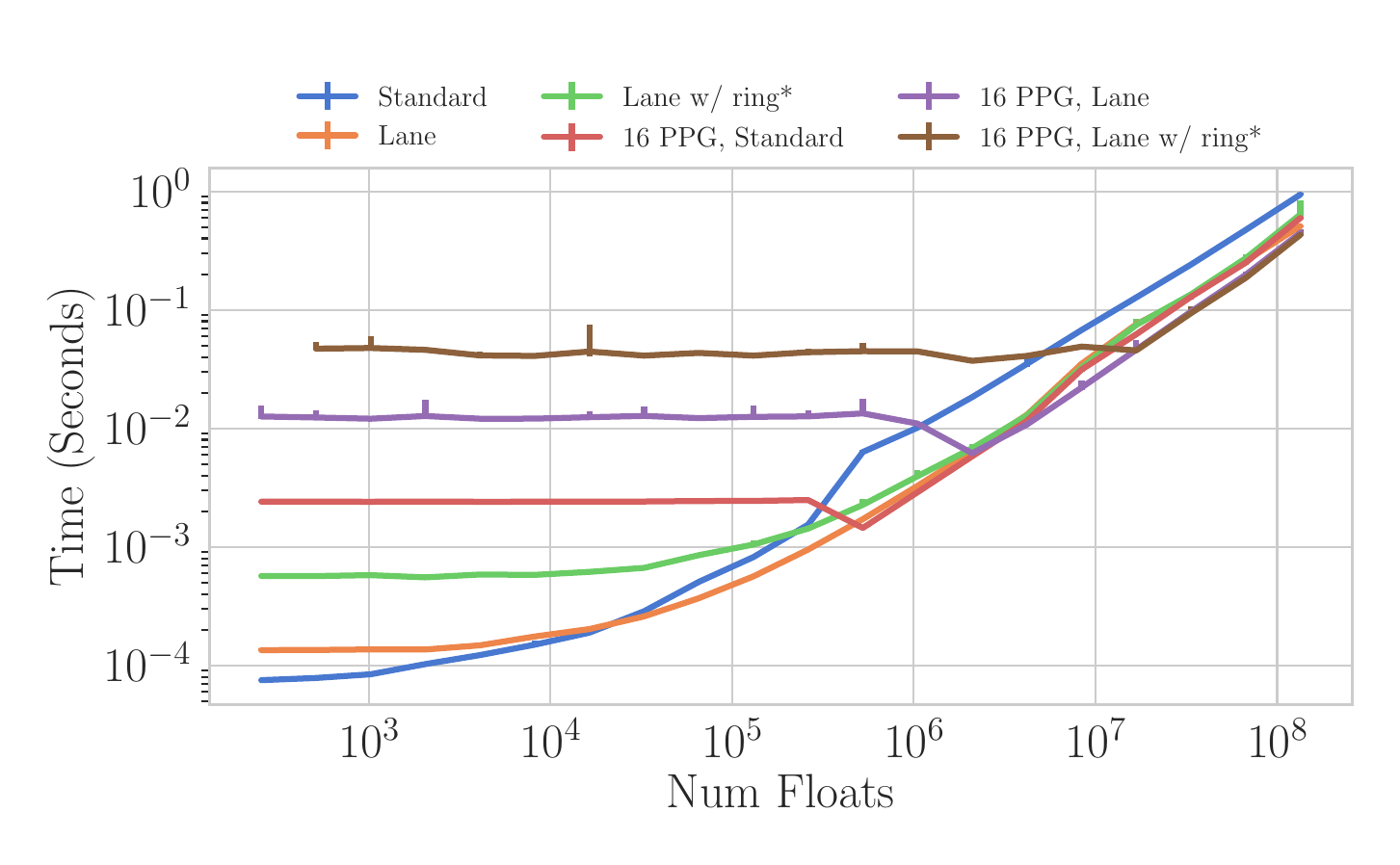}
    \caption{Standard and multi-lane algorithms with multiple processes per GPU on 8 nodes of Delta, varying buffer sizes}\label{fig:full_mpi_comparison1}
\end{figure}

\begin{figure}[ht!]
    \centering
    \includegraphics[width=\linewidth,page=11]{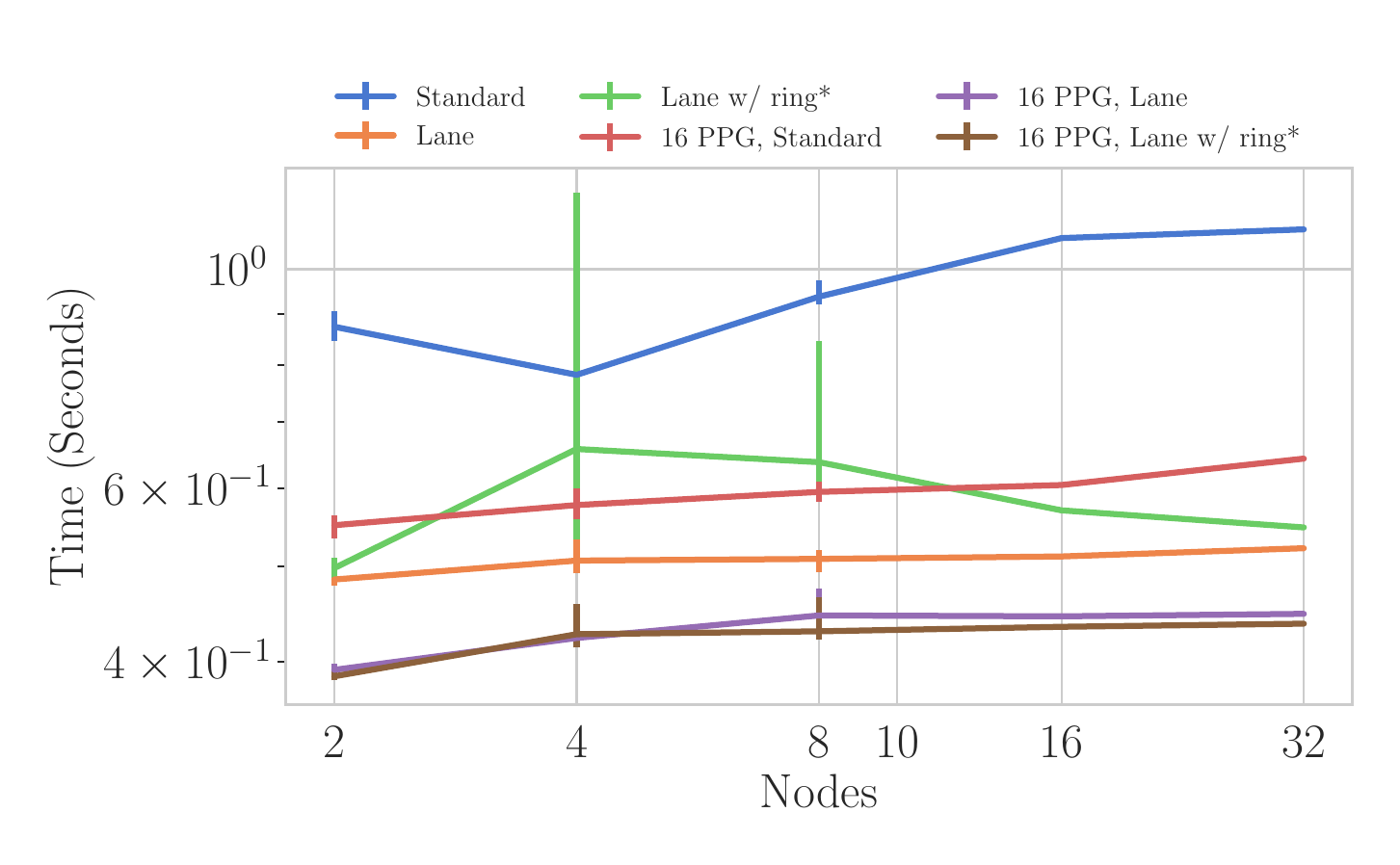}
    \caption{Standard and multi-lane algorithms with multiple processes per GPU of 134 million floats on Delta, varying node counts}\label{fig:full_mpi_comparison2}
\end{figure}

Finally, Figures~\ref{fig:full_mpi_comparison1} and \ref{fig:full_mpi_comparison2} compare the performance of all algorithms, including where the ring algorithm is used in the inter-node stage of the multi-lane all-reduce, comparing $1$ versus $16$ processes per GPU.  The multi-lane algorithm with $16$ processes per GPU achieves up to $2.4$x speedup over the standard \texttt{MPI\_Allreduce} for large data sizes. These results agree with our ping-pong microbenchmark on Delta.

\subsection{GPUDirect RDMA Analysis}\label{sec:gpudirect_results}

\begin{comment}
\begin{figure*}[ht!]
    \centering
    \begin{subfigure}{0.49\textwidth}
        \centering
        \includegraphics[width=\textwidth,page=1]{Figures/allreduce_plus_copy_n32_node_run_errorbar.pdf}
        \caption{32 nodes, varying buffer sizes}
    \end{subfigure}
    \hfill
    \begin{subfigure}{0.49\textwidth}
        \centering
        \includegraphics[width=\linewidth,page=1]{Figures/allreduce_plus_copy_all_nodes_run_errorbar.pdf}
    \caption{$2^{29}$ floats, varying node counts}
    \end{subfigure}
    \caption{Standard algorithm with multiple processes per GPU on Tuolumne}\label{fig:tuolumne}
\end{figure*}
\end{comment}

\begin{comment}
\begin{figure*}[ht!]
    \centering
    \begin{subfigure}{0.49\textwidth}
        \centering
        \includegraphics[width=\textwidth,page=15]{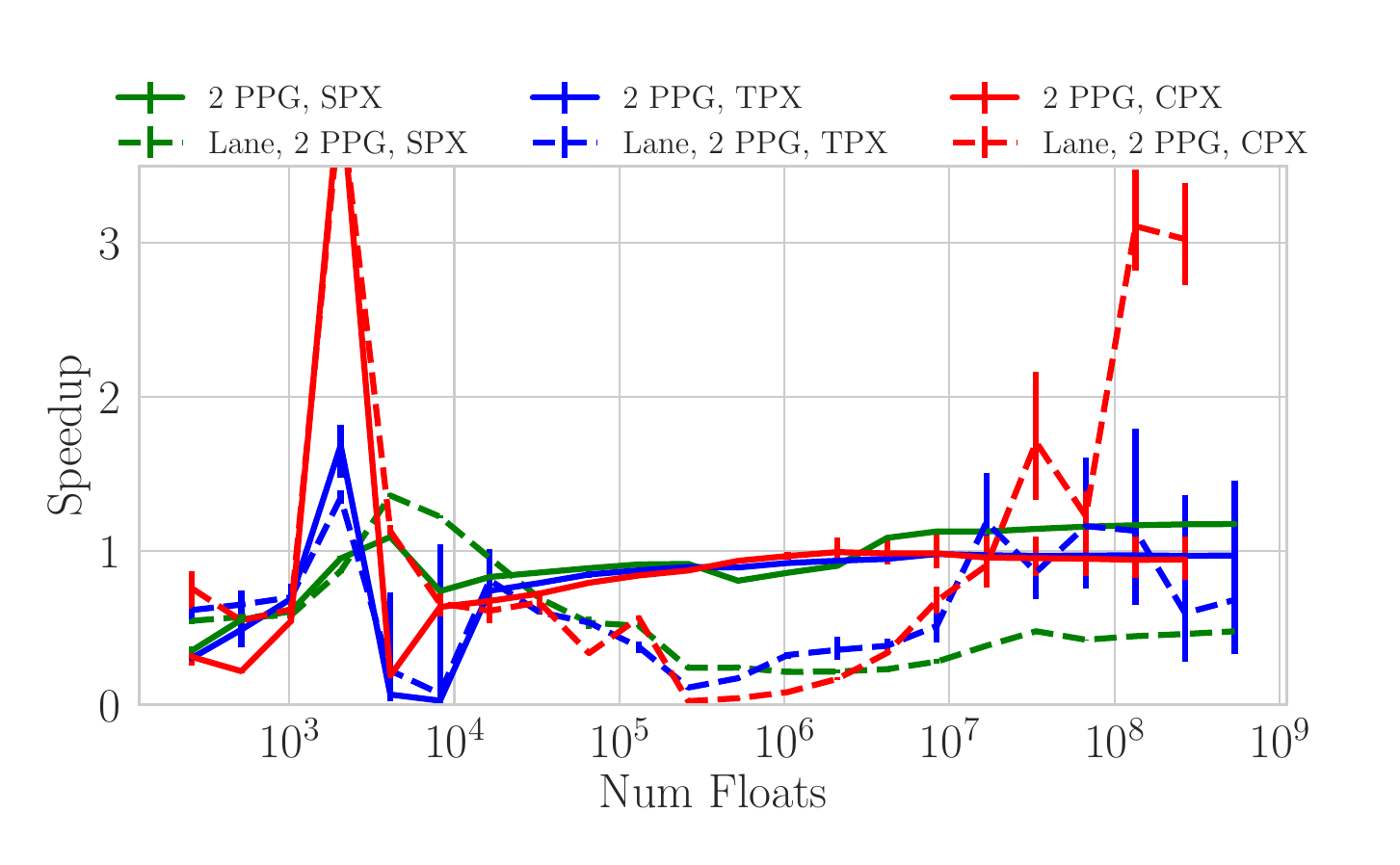}
        \caption{32 nodes, varying buffer sizes}\label{fig:spx_tpx_cpx_floats}
    \end{subfigure}
    \hfill
    \begin{subfigure}{0.49\textwidth}
        \centering
        \includegraphics[width=\linewidth,page=15]{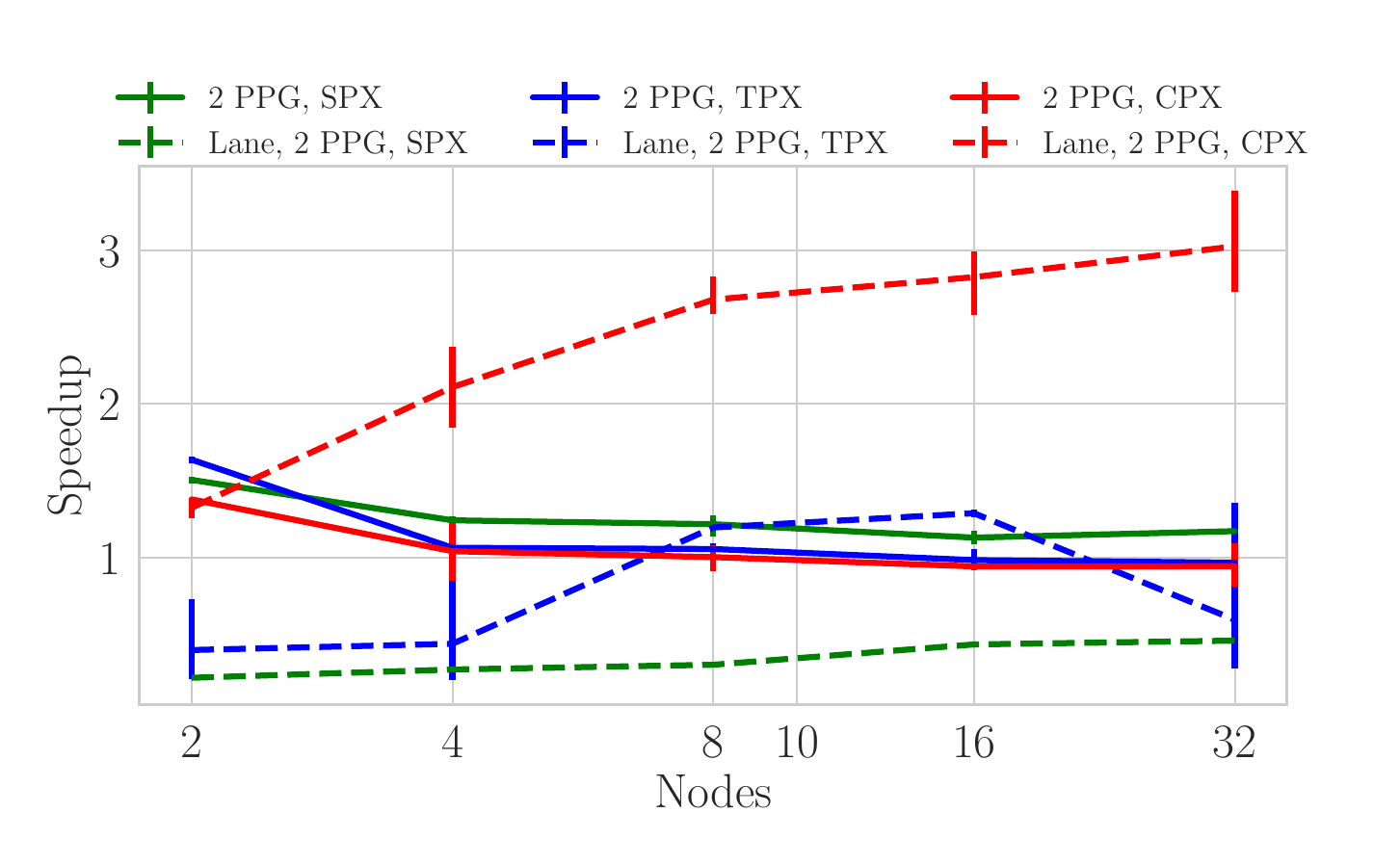}
    \caption{$2^{28}$ floats, varying node counts}\label{fig:spx_tpx_cpx_nodes}
    \end{subfigure}
    \caption{Multi-process standard and multi-lane algorithms on the SPX/TPX/CPX modes of the APU on Tuolumne}\label{fig:spx_tpx_cpx}
\end{figure*}
\end{comment}

The performance of GPUDirect RDMA-enabled MPI is analyzed on one of the fastest AMD architectures, Tuolumne, using Cray MPICH 9.0.1 and ROCm 6.4.0. For benchmarks using more than one process per GPU, data to be communicated was copied to and from an allocated device buffer before and after MPI communication because Cray MPICH does not support buffers shared via ROCm IPC. Each process performs an independent all-reduce operation on their assigned subset of the buffer.  All processes were assigned to the GPU adjacent to their NUMA region. Each benchmark was run 3 separate times; error bars in plots show run to run variation.

\begin{figure}[ht!]
    \centering
    \includegraphics[width=\linewidth,page=1]{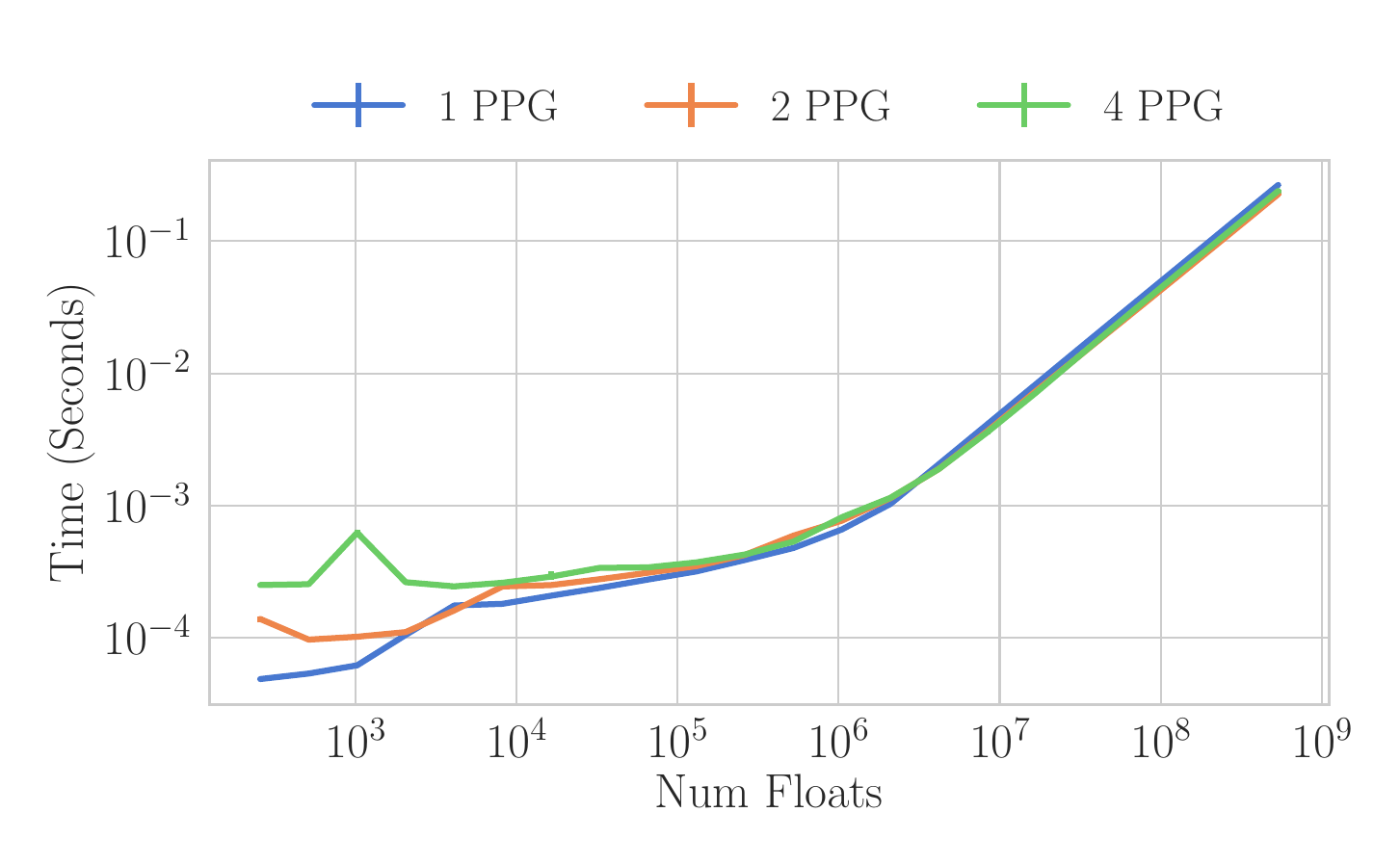}
    \caption{Standard algorithm with multiple processes per GPU on 32 nodes of Tuolumne, varying buffer sizes}\label{fig:tuolumne1}
\end{figure}

\begin{figure}[ht!]
    \centering
    \includegraphics[width=\linewidth,page=1]{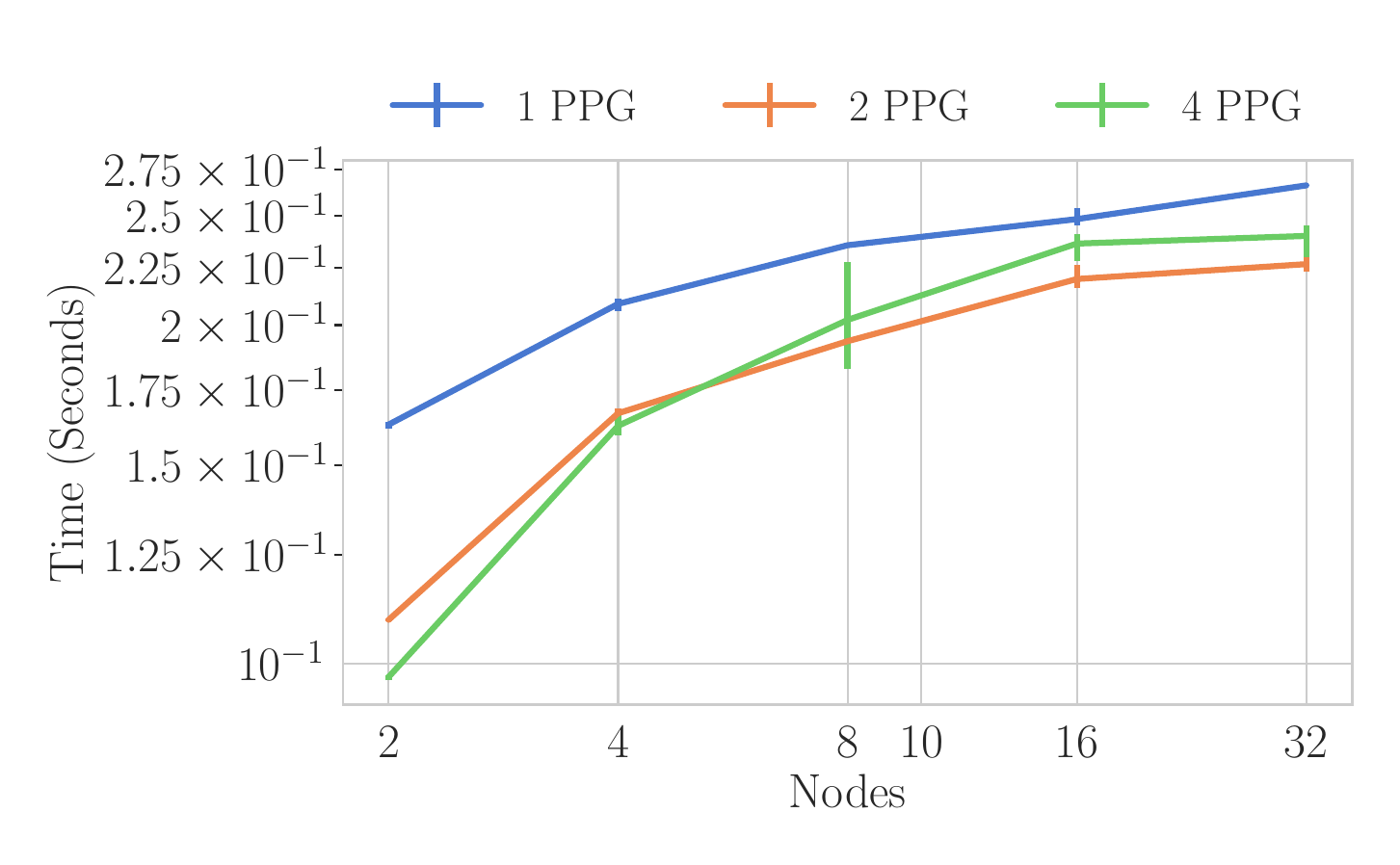}
    \caption{Standard algorithm with multiple processes per GPU of 536 million floats on Tuolumne, varying node counts}\label{fig:tuolumne2}
\end{figure}

Figures~\ref{fig:tuolumne1} and \ref{fig:tuolumne2} show the performance of an all-reduce when varying the number of processes per GPU from 1 to 4 on Tuolumne with each MI300A in SPX mode.  Figure~\ref{fig:tuolumne1} shows timings on 32 nodes at different sizes of the communication buffer where using multiple processes per GPU shows speedup for the largest buffer sizes. 

Figure~\ref{fig:tuolumne2} shows the timings of the largest buffer size as the number of nodes increases. We see steady speedup as the number of nodes increases with 2 and 4 processes per GPU, yielding speedups of $1.17$x and $1.1$x at 32 nodes, respectively. These speedups begin to decrease at large scales; more significant speedups of up to $1.21$x are seen at 8 nodes. Using more than 4 processes per GPU does not show any benefit on Tuolumne.

It is noted that a small benchmark was performed on 8 nodes of Tuolumne with the System Direct Memory Access (SDMA) engines of the MI300A both enabled and disabled.  Consistent with previous findings regarding small-scale inter-APU communication \cite{schieffer2025interapucommunicationamdmi300a}, the use of SDMA engines does not appear to impact the advantages provided by using multiple processes per GPU.

\subsubsection{MI300A Partitioning Modes}\label{sec:mi300a_partitioning}

\begin{figure}[ht!]
    \centering
    \includegraphics[width=\linewidth,page=15]{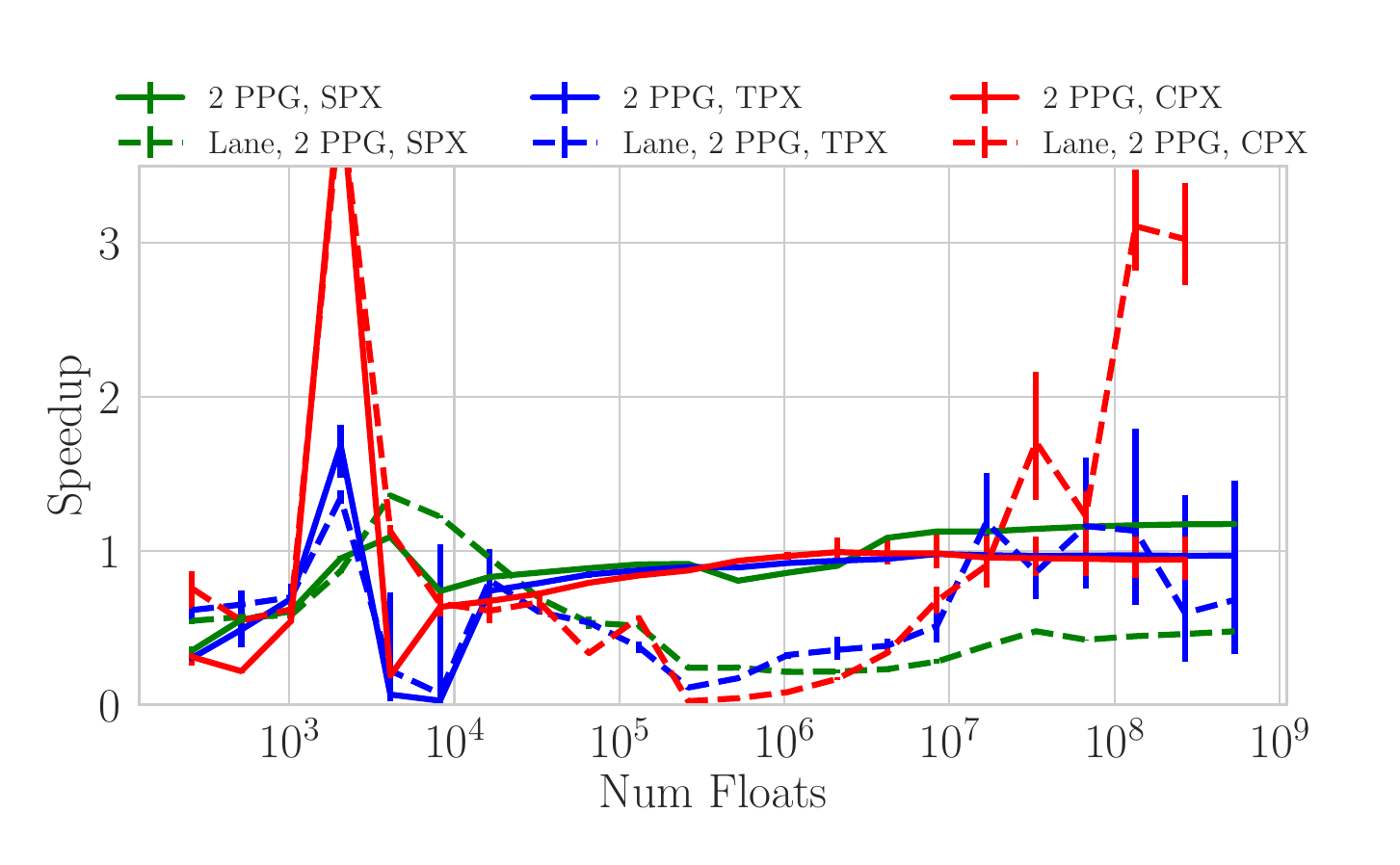}
    \caption{Standard and multi-lane algorithms with multiple processes per GPU on 32 nodes across the SPX/TPX/CPX modes of Tuolumne, varying buffer sizes}\label{fig:spx_tpx_cpx_floats}
\end{figure}

\begin{figure}[ht!]
    \centering
    \includegraphics[width=\linewidth,page=15]{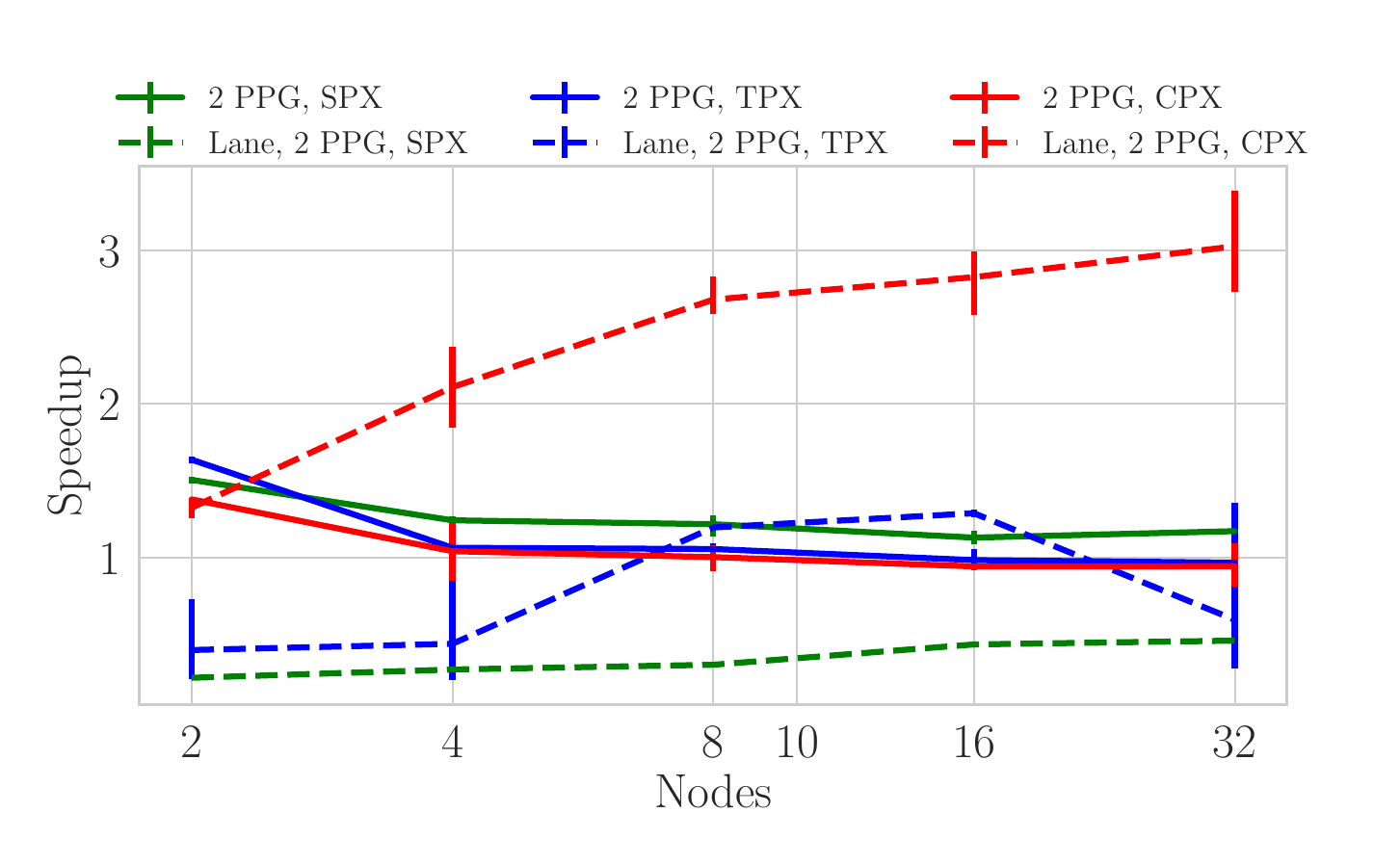}
    \caption{Standard and multi-lane algorithms with multiple processes per GPU of 268 million floats across the SPX/TPX/CPX modes of Tuolumne, varying node counts}\label{fig:spx_tpx_cpx_nodes}
\end{figure}

The AMD MI300A APUs were also tested in the TPX and CPX partitioning modes. Figures~\ref{fig:spx_tpx_cpx_floats} and \ref{fig:spx_tpx_cpx_nodes} show the speedups of using two processes per GPU for both proposed algorithms against the standard \texttt{MPI\_Allreduce} implementation for each partitioning mode~\footnote{It is important to note that speedups from each partitioning mode cannot be directly compared since the buffer size per logical GPU is not scaled with the number of logical GPUs per node.}. Figure~\ref{fig:spx_tpx_cpx_floats} shows speedups on 32 nodes of Tuolumne with different buffer sizes. In SPX and CPX modes with either the standand or multi-lane algorithm, respectively, there is a significant advantage to using two processes per GPU for larger buffers.

%Figure~\ref{fig:spx_tpx_cpx_nodes} shows speedups of both algorithms while the number of nodes increases at the largest buffer size tested across all partitioning modes. One of the novel algorithms presented provides speedup in SPX and CPX modes; we see that the multi-lane multi-process algorithm yields speedups of up to $3$x over the standard algorithm in CPX mode while the standard multi-process algorithm shows no speedup. Performance in TPX mode does not show benefit at scale for either algorithm. The standard multi-process algorithm is only advantageous in SPX mode. Thus, the multi-lane algorithm benefits from more logical GPUs potentially due to the increased memory locality near each XCD die~\cite{mi300a_roofline} with smaller communication sizes from the initial call to reduce-scatter. The standard algorithm exhibits the opposite, potentially benefiting from less logical GPUs per node, in agreement with the SPX ping-pong microbenchmark on Tuolumne.
%\enlargethispage{3.5\baselineskip}
Figure~\ref{fig:spx_tpx_cpx_nodes} shows speedups of both algorithms while the number of nodes increases at the largest buffer size tested across all partitioning modes. One of the novel algorithms presented provides speedup in SPX and CPX modes; we see that the multi-lane multi-process algorithm yields speedups of up to $3$x over the standard algorithm in CPX mode while the standard multi-process algorithm shows no speedup. Performance in TPX mode does not show benefit at scale for either algorithm. The standard multi-process algorithm is the only advantageous implementation in SPX mode. Thus, the multi-lane algorithm benefits from more logical GPUs. Working in tandem with the smaller communication sizes (a $6$x decrease from SPX to CPX mode), each intra-node group communicator, $comm_{group}$, has $4$ clusters of $6$ ranks residing within the same NUMA region, potentially accelerating the multi-lane intra-node steps in CPX mode whereas in SPX mode, all ranks in $comm_{group}$ are from separate NUMA regions. Further, there is likely increased memory locality near each XCD die in CPX mode~\cite{mi300a_roofline}. The standard algorithm benefits from less logical GPUs per node, agreeing with the ping-pong microbenchmark on Tuolumne.

We also observe that our multi-lane algorithm using two processes per GPU in CPX mode scales significantly better than the standard algorithm in any of the partitioning modes; the speedup of our multi-lane algorithm continues to increase as the number of nodes increases.

\begin{figure}[ht!]
    \centering
    \includegraphics[width=\linewidth,page=13]{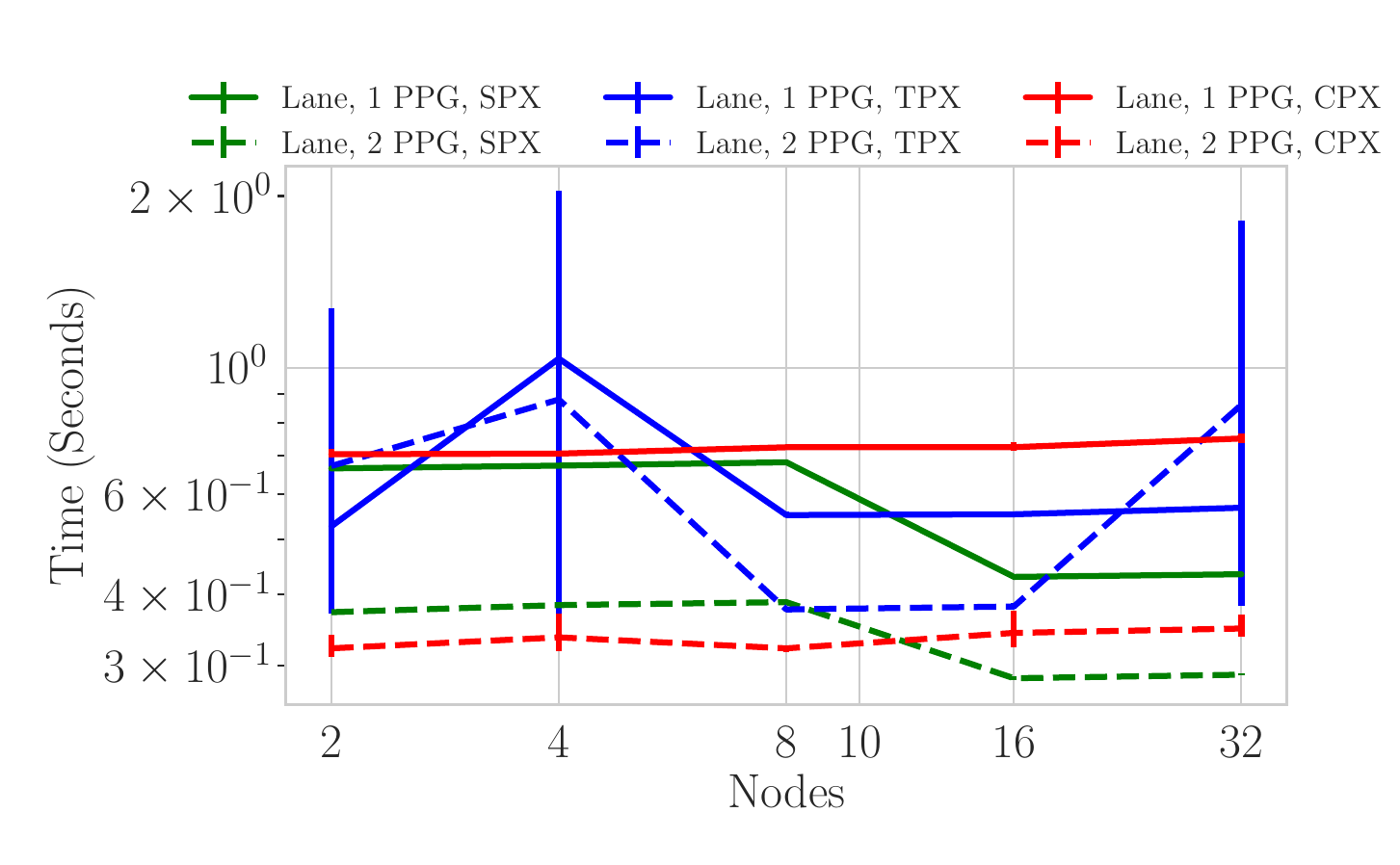}
    \caption{Multi-lane algorithm with multiple processes per GPU of 268 million floats across the SPX/TPX/CPX modes of Tuolumne, varying node counts}\label{fig:lane_spx_tpx_cpx}
\end{figure}
The performance of the multi-lane algorithm as the number of nodes increases can be seen in Figure~\ref{fig:lane_spx_tpx_cpx}.  We observe that using multiple processes per GPU benefits the multi-lane algorithm in SPX and CPX modes, yielding $1.5$x and $2.1$x speedup, respectively, emphasizing that the speedup seen in Figure~\ref{fig:spx_tpx_cpx_nodes} is due to the structure of the multi-lane algorithm, the use of multiple processes per GPU, and the fine-grained parallelism offered by CPX partitioning. 

\section{Conclusions and Future Work}\label{sec:conclusions}
\begin{comment}
\begin{itemize}
    \item Point out notable results and key takeaways
    \item Future work: at very large scales, may want to further partition steps (instead of comm with n-nodes, could take the square of n-nodes and first reduce in one dimension and then in the next).
    \item Emerging systems have APUs, so GPU and CPU memory is combined, allowing for CPUs to potentially be involved in collectives alongside the GPUs.
    \item General approach can be applied to other collective operations.
\end{itemize}
\end{comment}
%\enlargethispage{3.6\baselineskip}
In this paper, we proposed to utilize more of the available CPU cores across all GPUs on a node to speed up GPU-aware MPI collectives. Using an Ampere-based system (Delta) on OpenMPI backed by host copy communication with our standard and lane-aware approaches with the all-reduce, we showed up to an $2.45$x speedup on large buffers.  Using Cray MPICH backed by GPUDirect RDMA communication with our standard approach on an El Capitan-based system (Tuolumne), we showed a $1.17$x speedup. Across the three partitioning modes of the AMD MI300A on Tuolumne, we showed that the multi-process multi-lane all-reduce is highly performant in CPX mode, yielding a $3$x speedup at scale.

We propose to extensively evaluate the performance characteristics of our multi-process approaches and also extend them to collectives such as the neighborhood all-to-all used in sparse linear systems. Due to the hierarchical structure of modern clusters, future work may also extend this approach to cluster topologies, i.e. stages of reduction in fat tree networks, in order to reduce the number of processes required at each stage of the communication.  

Emerging systems like Tuolumne feature memory architectures that are shared between the CPU and GPU.  Future work could potentially examine tradeoffs in scientific applications by driving communication with more CPU cores while using GPUs for local computation, allowing these steps to overlap while minimizing synchronization costs.

\section*{Acknowledgments}
This work was performed with partial support from the National Science Foundation under Grant No. CCF-2338077 and CNS-2450092 and the U.S. Department of Energy's National Nuclear Security Administration (NNSA) under the Predictive Science Academic Alliance Program (PSAAP III \& IV) Awards DE-NA0003966 and DE-NA0004267. 

Any opinions, findings, and conclusions or recommendations expressed in this material are those of the authors and do not necessarily reflect the views of the National Science Foundation and the U.S. Department of Energy's National Nuclear Security Administration.

This research used the Tuolumne supercomputer at Lawrence Livermore National Laboratory.  This research used the Delta advanced computing and data resource which is supported by the National Science Foundation (award OAC 2005572) and the State of Illinois. Delta is a joint effort of the University of Illinois Urbana-Champaign and its National Center for Supercomputing Applications.

Generative AI was used to improve formatting of plots.

\begingroup
\begin{spacing}{0.88}
%\addtolength{\textheight}{3.5\baselineskip}
\bibliographystyle{IEEEtran}
\bibliography{refs}
\end{spacing}
\endgroup

\end{document}